\documentclass[a4paper,11pt]{article}
\usepackage{authblk}
\usepackage[british]{babel}
\usepackage[mono=false]{libertine}
\usepackage{setspace}

\usepackage[utf8]{inputenc} % allow utf-8 input
\usepackage[T1]{fontenc}    % use 8-bit T1 fonts
\usepackage{url}            % simple URL typesetting
\usepackage{booktabs}       % professional-quality tables
\usepackage{amsfonts}       % blackboard math symbols
\usepackage{nicefrac}       % compact symbols for 1/2, etc.
\usepackage{microtype}      % microtypography
\usepackage{mathtools}
\usepackage{amsmath,amsfonts,amssymb,amsthm,mathrsfs,bbm}
\usepackage{latexsym,amscd,amsbsy,dsfont}
\usepackage{mathtools}
\usepackage{dsfont}
\usepackage{color}
\usepackage[usenames, dvipsnames]{xcolor}
\usepackage{graphicx}
\usepackage{epstopdf}
\usepackage{float}
\usepackage{cases}
\usepackage[top=2.5cm, bottom=2.5cm, left=2cm, right=2cm]{geometry}
\usepackage{caption}
\usepackage{physics}
\usepackage{tensor}
\usepackage{subcaption}
\usepackage[most]{tcolorbox}
\tcbuselibrary{theorems}
\usepackage{empheq}
\usepackage{times}
\usepackage{stmaryrd}

\usepackage{xr-hyper} % or use \usepackage{xr}
\usepackage[pdfpagemode=UseNone,bookmarksopen=false,colorlinks=true,urlcolor=RoyalBlue,citecolor=YellowOrange,citebordercolor=RoyalBlue,linkcolor=RoyalBlue]{hyperref}
\usepackage{bookmark}

\usepackage{times}
\usepackage{enumitem}

\author{Alia Abbara$^\star$, Benjamin Aubin$^\dagger$, Florent Krzakala$^\star$, Lenka Zdeborov{\'a}$^\dagger$}
\date{
$\dagger$ \textit{Institut de Physique Th\'eorique \\
CNRS \& CEA \& Universit\'e Paris-Saclay, Saclay, France}\\ \vspace{0.3cm}
$\star$ \textit{Laboratoire de Physique Statistique\\
CNRS \& Sorbonnes Universit\'es \& \\
\'Ecole Normale Sup\'erieure, PSL University, Paris, France}\\
\vspace{1cm}
}

\title{Rademacher complexity and spin glasses:\\ A link between the replica and statistical theories of learning}

\renewcommand{\subsubsection}[1]{\textbf{#1}\\}

\def \({\left(}
\def \){\right)}
\def \[{\left[}
\def \]{\right]}

\newcommand{\bx}{{\textbf {x}}}

\newcommand{\btheta}{{\boldsymbol{\theta}}}

\newcommand{\bepsilon}{{\boldsymbol{\epsilon}}}

\renewcommand{\d}{\text{d}}

\newcommand{\sign}{\text{ sign}}

\newcommand{\be}{\begin{equation}}
\newcommand{\ee}{\end{equation}}
\newcommand{\beqa}{\begin{eqnarray}}
\newcommand{\eeqa}{\end{eqnarray}}

\newcommand{\bea}{\begin{align}}
\newcommand{\eea}{\end{align}}

\newtheorem{theorem}{Theorem}[section]

\newtheorem{definition}[theorem]{\textbf{Definition}}

\DeclareMathAlphabet{\varmathbb}{U}{bbold}{m}{n}
\newcommand{\id}{\mathds{1}}
\newcommand{\EE}{\mathbb{E}}
\newcommand{\bbR}{\mathbb{R}}
\newcommand{\bbP}{\mathbb{P}}

\newcommand{\bbN}{\mathbb{N}}

\renewcommand{\d}{{\rm d}}

\newcommand{\mZ}{\mathcal{Z}}
\newcommand{\mH}{\mathcal{H}}
\newcommand{\mO}{\mathcal{O}}
\newcommand{\mI}{\mathcal{I}}

\newcommand{\mN}{\mathcal{N}}

\newcommand{\mC}{\mathcal{C}}

\newcommand{\mL}{\mathcal{L}}

\newcommand{\mR}{\mathcal{R}}
\newcommand{\mF}{\mathcal{F}}

\newcommand{\extr}{\textrm{\textbf{extr}}}
\renewcommand{\tr}[1]{\textrm{Tr}\[#1\]}
\newcommand{\td}[1]{{\tilde{#1}}}

\newcommand{\spacecase}[0]{\vspace{0.3cm} \\}

\renewcommand{\subparagraph}[1]{\vspace{0.5cm} $\bullet$ \textit{#1} \\ \vspace{0.2cm} }
\newcommand{\andcase}[0]{\hspace{ 0.2cm }\textrm{ and }\hspace{ 0.2cm }}
\newcommand*\diff{\mathop{}\!\mathrm{d}}
\newcommand*\Diff{\mathop{}\!\mathrm{D}}

\renewcommand{\vec}[1]{{\textbf{#1}}}
\newcommand{\mat}[1]{{\rm #1}}
\newcommand{\tbf}[1]{{\bold{#1}}}

\newcommand{\iid}[0]{{\rm i.i.d }}

\begin{document}
\maketitle

\begin{abstract}
Statistical learning theory provides bounds of the generalization gap, using in particular the Vapnik-Chervonenkis dimension and the Rademacher complexity. An alternative approach, mainly studied in the statistical physics literature, is the study of generalization in simple synthetic-data models. Here we discuss the connections between these approaches and focus on the link between the Rademacher complexity in statistical learning and the theories of generalization for \emph{typical-case} synthetic models from statistical physics, involving quantities known as \emph{Gardner capacity} and \emph{ground state energy}. We show that in these models the Rademacher complexity is closely related to the ground state energy computed by replica theories. Using this connection, one may reinterpret many results of the literature as rigorous Rademacher bounds in a variety of models in the high-dimensional statistics limit. Somewhat surprisingly, we also show that statistical learning theory provides predictions for the behavior of the ground-state energies in some full replica symmetry breaking models.
\end{abstract}

%\setcounter{tocdepth}{1}
%\renewcommand\contentsname{}
%\tableofcontents

%%% General Introduction %%%
\section{\label{sec:introduction}Introduction}
Empirical risk minimization is the workhorse of most of modern supervised machine learning successes. Consider for instance a data-set $\left\{y^{(\mu)} ,\vec{x}^{(\mu)} \right\}_{\mu=1}^m$ of $m$ examples ${\bx}^{(\mu)} \in \mathbb{R}^d$ assumed to be drawn from a distribution $P_x(.)$, with labels $y^{(\mu)} \in \{ -1, +1 \}$ used for a binary classification task. We consider an estimator $f_{\vec{w}}(.)$ that belongs to a \emph{hypothesis class} $\mF$, for instance a neural network or a linear function, with respective weights or parameters $\vec w$. The latter are typically computed by minimizing the empirical risk
\begin{equation}
    {\mR}_{\rm empirical}^m ( f_{\vec{w}} ) = \frac{1}{m} \sum_{\mu=1}^m {\mL}\left( y^{(\mu)} ,f_{\vec w} \left( \vec{x}^{(\mu)}\right) \right)  \nonumber \,
\end{equation}
over $\vec{w}$, where $\mL$ denotes a loss function, e.g. the mean-squared-loss ${\cal L}(a,b)=(a-b)^2$. The main theoretical issue of statistical learning theory concerns the performance of the estimator $f_{\vec{w}}(.)$ obtained by such a minimization on yet unseen data, namely the \emph{generalization problem}. In fact, what we really hope to minimize is the population risk, defined as
\begin{equation}
    {\mR}_{\rm population} (f_{\vec{w}}) = 
    \mathbb{E}_{y,{\textbf x}} \left[{\cal L}(y ,f_{\vec{w}}({\vec{x})})\right]  \nonumber \,.
\end{equation}
Since we are optimizing the empirical risk instead, the difference between the two might be arbitrarily large. Bounding this difference between \emph{empirical} and \emph{population} risks is therefore a major problem of statistical learning theories.

In a large part of the literature, statistical learning analysis (see e.g.~\cite{bartlett2002rademacher,vapnik2013nature,shalev2014understanding}) relies on the Vapnik-Chervonenkis (VC) analysis and on the so-called \emph{Rademacher complexity}. The latter is a measure of the complexity of $\mF$,  the hypothesis class spanned by $f_{\vec{w}}(.)$, to bound ${{\mR}_{\rm population} - {\cal R}^m_{\rm empirical}}$, the \emph{generalization gap}. A gem within the literature is the Uniform Convergence result which states the following: if the Rademacher complexity or the VC dimension is finite, then for a large enough number of samples the generalization gap will vanish uniformly over all possible values of parameters ${\vec w}$. Informally, uniform convergence tells us that with high probability, for any weights value $\vec{w}$, the generalization gap satisfies
\vspace{-0.25cm}
\begin{equation}
  {\mR}_{\rm population}(f_{\vec{w}}) - {\cal R}_{\rm empirical}^m(f_{\vec{w}}) = \mO\left(\sqrt{\frac{d_{\rm VC}(\mF)}{m}}\right)\,,
  \label{eq:VC-bound}
  \vspace{-0.2cm}
\end{equation}
where $d_{\rm VC}(\mF)$ denotes the Vapnik-Chervonenkis dimension of the hypothesis class $\mF$. Tighter bounds can be obtained using the Rademacher complexity. These  bounds, although useful, do not seem to fully explain the success of current deep-learning architectures (\cite{zhang2016understanding}).

Over the last four decades, a different vision of generalization  --- based on the analysis of {\it typical case} problems with synthetic data created from simple generative models --- was developed to a large extent in the statistical physics literature (see e.g. \cite{Seung1992,Watkin1993,opper1995statistical,engel2001statistical} for a review). The link with the VC dimension was discussed in many of these works, notably via its connection with its  twin from statistical physics, the \emph{Gardner capacity} (\cite{gardner1988optimal}). In particular, one can show that the VC capacity is always larger than half of the Gardner one (\cite{engel2001statistical}). We shall review this discussion later on in this paper. 
However, to the best of our knowledge the Rademacher complexity was absent from these considerations. This omission is unfortunate: not only does the Rademacher complexity give tighter bounds than the VC dimension, it also intrinsically connects with a quantity that physicists are familiar with and have been computing from the very beginning of their studies, namely the average \emph{ground-state energy}. 

The goal of the present paper is to bridge this gap and unveil the deep link between ground-state energy and Rademacher complexity, and how this connection is valuable to both parties. The paper is organized as follows: After giving proper definitions of common generalization bounds in sec.~\ref{sec:rademacher}, we detail calculations of Rademacher complexities for simple function classes in sec.~\ref{sec:iid}. These sections serve as an introduction to the readers not familiar with these notions.  The subsequent sections~\ref{sec:statistical physics} and \ref{sec:applications} provide the original content of the paper.
\paragraph{Here we summarize the main contributions of this paper:}
\begin{itemize}[topsep=1pt,itemsep=1pt,partopsep=1pt, parsep=1pt]
 \setlength\itemsep{0.1em}
    \item  We point out the one-to-one connections between the Rademacher complexity in statistical learning, and the ground-state energies and Gardner capacity from statistical physics.
    \item We show how the heuristic replica method from statistical physics can be used to compute the Rademacher complexity in the high-dimensional statistics limit and reinterpret classical results of the statistical physics literature as Rademacher bounds in the case of perceptron and committee machine models with \iid data.
    \item  We contrast these results with the generalization in the teacher-student scenario, illustrating the worst-case nature of the Rademacher bound that fails to capture the typical-case behavior.
    \item We finally show {\it en passant}, that learning theory also bears consequences for the spin glass physics and the related replica symmetry breaking scheme by showing it implies  strong constraint on the ground-state energy of some spin glass models.
\end{itemize}

%%% Handwaving explanation of the rademacher %%%
\section{A primer on Rademacher complexity \label{sec:rademacher}}

The bound of the generalization gap involving the VC dimension is specific to binary classification, and does not depend on the data distribution. While this is a strong property, the Rademacher approach does depend on data distribution and allows for tighter bounds. Moreover, it generalizes to multi-class classification and regression problems. We recall the definition of the Rademacher complexity:
\begin{definition}
	Let $f_{\vec{w}}$ be any function in the hypothesis class $\mF$, and let $\bepsilon \in \{ \pm 1 \}^m$ be drawn uniformly at random. The \textbf{empirical Rademacher complexity} is defined as
	\begin{align}
		\hat{\mathfrak{R}}_m \( \mF, \mat{X} \) \equiv \EE_{\bepsilon} \[ \sup_{f_{\vec{w}} \in \mF} \frac{1}{m} \sum_{\mu =1}^m  \epsilon_\mu  f_{\vec{w}}\(\vec{x}^{(\mu)}\) \] \,,
		\label{main:def_empirical_rademacher}
	\end{align} 
	and depends on the sample examples $\mat{X} = \{\vec{x}^{(1)}, \dots \vec{x}^{(m)} \} \in \bbR^{d \times m}$. The \textbf{Rademacher complexity} is defined as the population average
	\begin{align}
		\mathfrak{R}_m \( \mF \) \equiv \EE_{\mat{X}} \[\hat{\mathfrak{R}}_m \( \mF, \mat{X} \) \] \,.
		\label{main:def_rademacher}
	\end{align} 
\end{definition}

In this paper, we shall focus on binary classification and consider the corresponding loss function $\mL(a,b) = \mathbbm{1} \[ a \ne b \]$ that counts the number of misclassified samples. We will be therefore interested in a hypothesis class $\mF =\left \{ f_{\vec{w}}: \bbR^d \mapsto \{\pm 1 \} \right\}$. Defining the training $\epsilon_{\rm train}^m(.)$ and generalization errors $\epsilon_{\rm gen}(.)$ for any function $f_{\vec{w}}\in \mF$ by
\begin{align}
    \epsilon_{\rm train}^m (f_{\vec{w}}) & \equiv \frac{1}{m} \sum_{\mu=1}^{m} \id \[ y^{(\mu)} \ne f_{\vec{w}}\(\vec{x}^{(\mu)}\) \] \andcase \epsilon_{\rm gen} (f_{\vec{w}}) \equiv \EE_{y,\vec{x}} \[ \id \[ y \ne f_{\vec{w}}\(\vec{x}\) \] \]\,,
\end{align}
the Rademacher complexity  provides a generalization error bound as expressed by the following theorem, and many of its variants (see e.g. \cite{bartlett2002rademacher,vapnik2013nature,shalev2014understanding,Mohri2018}):
\begin{theorem}[Uniform convergence bound - Binary classification]
Fix a distribution $P_x$ and let $\delta > 0$. Let $\mat{X}= \{\vec{x}^{(1)}, \dots \vec{x}^{(m)} \} \in \bbR^{d \times m}$ be drawn \iid from $P_x$. Then
with probability at least $1 - \delta$ (over the draw of $\mat{X}$),
\begin{align}
      \forall f_{\vec{w}} \in \mF,~~ \epsilon_{\rm gen} (f_{\vec{w}}) - \epsilon_{\rm train}^m (f_{\vec{w}})   \leq  \mathfrak{R}_m(\mF) + \sqrt{ \frac{\log(1/\delta)}{m}}  \,.
      \label{eq:main-bound}
    \end{align}
\end{theorem}
Thus, the Rademacher complexity is a uniform bound of the generalization gap. In the high-dimensional limit, i.e when both $m$ and $d$ go to infinity, that we will consider in the remaining of the paper we shall see that we
can discard the $\delta-$dependent term and that only the first term will remain finite. 

Note that this theorem can be used to recover the classical result \eqref{eq:VC-bound}. Indeed it can be shown (\cite{massart2000some,ledoux2013probability,dudley1967sizes}) that the Rademacher complexity can be bounded by the VC dimension so that for some constant value $C$,
\begin{equation}
\mathfrak{R}_m(\mF) \le  C{\sqrt {\frac {d_{\rm VC}(\mF)}{m}}}  \,.
\label{boundVC}
\end{equation}
We remind the reader that the VC dimension is the size of the {\rm set} that can be fully shattered by the hypothesis class $\mF$. Informally, if $m > d_{\rm VC}$ then for all set of $m$ data points, there exists an assignment of labels that cannot be fully fitted by the function class (\cite{vapnik2013nature}).

%%% Handwaving explanation of the rademacher %%%
\section{Synthetic models in the high-dimensional statistics limit}
\label{sec:iid}
In this section, we consider data generated by a simple generative model. We suppose that each vector of the input data points $\mat{X}=\{\vec{x}^{(1)}, \cdots, \vec{x}^{(m)}\} \in \bbR^{d \times m}$ has been generated \iid from a factorized, e.g. Gaussian, distribution, that is
$\forall \mu \in \llbracket 1;m \rrbracket, {P_x\left(\vec{x}^{(\mu)}\right) = \prod_{i=1}^{d} P_x(x_i^{(\mu)})}$.
In the following, we will focus on this simple data distribution, but sec.~\ref{sec:rot_inv_mat} presents a generalization to rotationally invariant data matrices $\mat{X}$ with arbitrary spectrum. The main interest of such settings is to use the analysis of \emph{typical case} problems with synthetic data created from simple generative models as means of getting additional insight on real world applications where data are not worst case (\cite{Seung1992,Watkin1993,opper1995statistical,engel2001statistical,Zdeborova2016}). In particular, we shall be interested in the high-dimensional statistics limit when $m, d \longrightarrow \infty$, with $\alpha=\frac{m}{d} = \Theta(1)$. In this paper, the aim is to compute exactly (rather than merely bounding) and asymptotically the Rademacher complexity for such problems.

\subsection{Linear model}
As the simplest example, we first tackle the computation of the Rademacher complexity for a simple function class containing all linear models with weights $\vec{w} \in \bbR^{d}$,
 \begin{align}
    \mF_{\rm linear} = \left\lbrace f_{\vec{w}}: 
\begin{cases}
\bbR^{d} \mapsto \bbR\\
\vec{x} \mapsto  \frac{1}{\sqrt{d}} \vec{w}^\intercal \vec{x}
\end{cases}  , \vec{w} \in \bbR^{d}~~/~~\|\vec{w}\|_2 = \Gamma \sqrt{d} \right\rbrace \,. 
\label{main:linear_model}
\end{align}
%$\hat{\vec{w}}$: $ f_{\vec{w}} : \mat{X} \longrightarrow  \frac{1}{\sqrt{d}} \mat{X}^\intercal \vec{w} \in \mF_{\rm linear}$  .
From eq.~\eqref{main:def_rademacher}, computing the empirical Rademacher complexity amounts to finding the vector $\vec{w}^\star$ that maximizes the scalar product between $\vec{y}$ (that replaces the variable $\bepsilon$) and $\mat{X}^\intercal \vec{w}$. It is thus sufficient to take $\vec{w}^\star = \frac{\mat{X} \vec{y}}{\|\mat{X} \vec{y}\|_2} \|\vec{w}\|_2$ and the empirical Rademacher complexity \eqref{main:def_rademacher} thus reads
\begin{align}
    \mathfrak{R}_m\left(\mF_{\rm linear}\right) &=\EE_{\vec{y}, \mat{X}} \left[  \frac{1}{m} \frac{1}{\sqrt{d}} \|\mat{X} \vec{y}\|_2 \|\vec{w}\|_2 \right]\,.
    %&=  \EE_{\vec{y}, \mat{X}} \[ \sup_{f \in \mF_{\rm linear}} \frac{1}{m} \sum_{\mu =1}^m  y^{(\mu)}  f\(\vec{x}^{(\mu)}\) \] = \EE_{\vec{y}, \mat{X}} \[ \frac{1}{m} \vec{y}^\intercal \( \frac{1}{\sqrt{d}} \mat{X}^\intercal\vec{w}^\star \) \] \\
    %&= \EE_{\vec{y}, \mat{X}} \[ \frac{1}{m} \vec{y}^\intercal \( \frac{1}{\sqrt{d}} \mat{X}^\intercal  \frac{\mat{X} \vec{y}}{\|\mat{X} \vec{y}\|_2} \|\vec{w}\|_2  \) \]  \|\vec{w}\|_2 \] \,.
\end{align}
$\mat{X}$ having \iid entries, we can apply the central limit theorem, which enforces $\forall i \in \llbracket 1,d \rrbracket$, $\left(\mat{X} \vec{y}\right)_i = \sum_{\mu=1}^m x_{i}^{(\mu)} y_\mu \sim \mN\left( 0 , m \right)$ hence ${\EE_{\vec{y}, \mat{X}} \|\mat{X} \vec{y}\|_2 = \sqrt{d m}}$. Assuming that weights are restricted to lie on the $\bbR^{d}$ sphere of radius $\Gamma$, we set $\|\vec{w}\|_2 = \Gamma \sqrt{d}$ and finally obtain
\begin{align}
     \mathfrak{R}_m\left(\mF_{\rm linear}\right) &= \frac{\Gamma}{\sqrt{\alpha}}\,\,,
\end{align}
where recall $\alpha = \frac{m}{d}$. The above result for the simple linear function hypothesis class allows to grasp the meaning of the Rademacher complexity: At fixed input dimension $d$, it decreases with the number of samples as $\alpha^{-1/2}$, closing the generalization gap in the infinite $\alpha$ limit. Illustrating the bias-variance trade-off, we also see that increasing the radius of the weights expands the function complexity (and might help for fitting the data-set), but unfortunately leads to a looser generalization bound.

Note also that the fact that the Rademacher complexity is $\Theta(\alpha^{-1/2})$ shows that it remains finite in the high-dimensional statistics limit. In this case, we see indeed that we can disregard the term $\sqrt{\log(1/\delta)/m}$ that goes to zero as $m\!\to\!\infty$ in eq.~(\ref{eq:main-bound}).

\subsection{Perceptron model}
The scaling of the Rademacher complexity inverse as $\sqrt{\alpha}$ in the high-dimensional statistics limit is actually {\it not} restricted to the linear model but appears to be a universal property, at least at large enough $\alpha$. To see this we now focus on a different hypothesis class: the perceptron, denoted $\mF_{\rm sign}$. This class contains linear classifiers which output binary variables, and will fit much better labels in the binary classification task. The class is defined as
 \begin{align}
    \mF_{\rm sign} = \left\lbrace f_{\vec{w}}: 
\begin{cases}
\bbR^{d} \mapsto \{\pm1\}\\
\vec{x} \mapsto  \sign\( \frac{1}{\sqrt{d}} \vec{w}^\intercal \vec{x} \)
\end{cases}  , \vec{w} \in \bbR^{d} \right\rbrace \,. 
\label{main:sign_model}
\end{align}
Let us consider a sample \iid matrix $\mat{X} \in \bbR^{d \times m}$ with $\vec{x}^{(\mu)}\sim \mN(\vec{0},\mat{I}_{d})$.
\begin{theorem} For the perceptron model class  eq.~\eqref{main:sign_model} with random i.i.d. input data in the high-dimensional limit, $\mathfrak{R}_m\left(\mF_{\rm sign}\right) = \Theta\left(\frac{1}{\sqrt{\alpha}} \right) $\,.
\end{theorem}
The proof is given in Appendix \ref{appendix:rademacher_linear_models}. In a nutshell, it uses the fact that  Rademacher complexity is upper-bounded by the VC dimension divided by $\alpha^{1/2}$, and lower-bounded by one particular example of its function class, when the weights are chosen according to Hebb's rule (\cite{hebb1962organization}), which also gives a behavior scaling as $\alpha^{-1/2}$. 

Heuristically, this result generalizes as well to a two-layer neural network with $K$ hidden neurons. Indeed, the two-layer function class contains, as a particular case, the single layer one, so the lower bounds goes through. The upper bound is however harder to control rigorously. Since neural networks have a finite VC dimension, the Rademacher complexity is again lower-bounded by $\mO(1/\sqrt{m})$; However, we do not know of any theorem that would ensure that the VC dimension is bounded by  $\mO(d)$ (\cite{bartlett2003vapnik}). Nevertheless, anticipating on the statistical physics approach, we indeed expect from the concentration (self-averaging) properties of the ground-state energy (\cite{talagrand2003spin}) in the high-dimensional limit that it will yield a Rademacher complexity that is a function of $\alpha=m/d$ {\it only} at fixed $K$. From this argument, we expect  that the $\Theta\left(\frac{1}{\sqrt{\alpha}} \right)$ dependence of the Rademacher complexity to be very generic in the high-dimensional limit.

%%% Different ways of computing it %%%
\section{The statistical physics approach\label{sec:statistical physics}}

\subsection{Average case problems: Statistical physics of learning}
As anticipated in the previous chapter, the approach inspired by statistical physics to understand neural networks considers a set of data points coming from known distributions. Again, for the purpose of this presentation we focus on a simple example, where $\vec x \sim P_{x}(.)$ with $P_x(\vec{x})= {\mN}_\vec{x}(\vec{0},\mat{I}_d)$. Sec.~\ref{sec:rot_inv_mat} is devoted to a generalization to random input data corresponding to random matrices with arbitrary singular value density.

Consider a function class, for instance we can again use the \emph{perceptron} one $\mF_{\rm sign}$: $\{f_{\vec{w}}: \vec{x} \to \left.\sign{\(\frac{1}{\sqrt{d}} \vec{w}^\intercal\vec{x}\)}\right\}$; a typical question in the literature was to compute how many misclassified examples can be obtained for a given rule used to generate the labels (\cite{engel2001statistical}). Given $m$ samples $\{y^{(\mu)}, \vec{x}^{(\mu)}  \}_{\mu=1}^m$, in order to count the number of wrongly classified training samples, we define the \emph{Hamiltonian}, or \emph{energy} function \cite{Mezard1986}:
\begin{align}
	\mH\(\{\vec{y},\mat{X}\}, \vec{w}\) &\equiv \sum_{\mu=1}^m \mathbbm{1} \[ y^{(\mu)} \ne 
    f_{\vec{w}}\( \vec{x}^{(\mu)}\) \] = \frac{1}{2}\(m - \sum_{\mu=1}^m   y^{(\mu)} f_{\vec{w}}\( \vec{x}^{(\mu)}\) \) \,.
\label{main:hamiltonian}
\end{align}
A classical problem in statistical physics is to compute the random capacity also called \emph{Gardner capacity} $\alpha_c$ (\cite{Gardner1989}): given $m$ examples $\{\vec{x}^{(\mu)}\}_{\mu=1}^{m}$ and labels  $\{y^{(\mu)}\}_{\mu=1}^{m}$ randomly chosen between $\pm 1$, it consists in finding how many samples $m_c$ can be correctly classified.

It turns out there exists a deep connection between the Gardner capacity and the VC dimension, as their common aim is to measure the maximum number of points $m_c$ such that there exists a function in the hypothesis class being able to fit the data set. In particular, using Sauer's lemma (\cite{SAUER1972145}) in the large size limit $m, d \longrightarrow \infty$, keeping $\alpha_{c}=\frac{m_c}{d} = \Theta(1)$ and $\alpha_{\rm VC} = \frac{d_{\rm VC}}{d} = \Theta(1)$, it is possible to show that the Gardner capacity $\alpha_c$ provides a lower-bound of the VC dimension (\cite{engel2001statistical}):
\begin{align}
    \alpha_c \leq 2  \alpha_{\rm VC} \,.
\end{align}{}
To illustrate this inequality, let us consider again the perceptron classifier hypothesis class $\mF_{\rm sign}$ for which the above inequality is saturated. In fact, the VC dimension is in this case (linear classification with binary outputs) simply $d_{\rm VC} = d$. Hence on one hand $\alpha_{\rm VC} = 1$ and on the other hand the Gardner capacity amounts to $\alpha_c=2$ (\cite{cover1965geometrical,Gardner1989}).

It is fair to say that a large part of the statistical physics literature focused mainly on the Gardner capacity, in particular in a series of works in the 90's  (\cite{Gardner1989, Krauth1989}) that led to more recent rigorous works (\cite{talagrand2003spin, talagrand2006parisi, Sun2018, Aubin2019}). 

\subsection{The Rademacher complexity and the ground-state energy}
As we shall see now, computing the Rademacher complexity for random input data can be directly reduced to a more natural object in the physics literature: the \emph{ground-state energy}. Defining the Gibbs measure at inverse temperature $\beta$, that weighs configurations with their respective cost, as
\begin{align}
 \langle   \dots \rangle_\beta  \equiv  \frac {\int d\vec{w} \dots e^{-\beta \mH(\{\vec{y},\mat{X}\}, \vec{w}) }}{\int d\vec w e^{-\beta \mH(\{\vec{y},\mat{X}\}, \vec{w})}}\, ,
\end{align}
we observe that averaging the Hamiltonian in eq.~\eqref{main:hamiltonian} over $\{\vec{y},\mat{X}\}$ and the Gibbs measure at temperature $\beta$ for any function $f_{\vec{w}} \in \mF$ provides
\begin{align}
   	\EE_{\vec{y}, \mat{X}} \left \langle  \frac{ 	\mH\(\{\vec{y},\mat{X}\}, \vec{w}\) }{d} \right\rangle_\beta &=  \frac{\alpha}{2} \[ 1 - \EE_{\vec{y}, \mat{X}} \left\langle \frac{1}{m} \sum_{\mu=1}^m y^{(\mu)}  f_{\vec{w}}\(\vec{x}^{(\mu)}\) \right\rangle_\beta   \] \,,
\end{align}
where $\alpha = \frac{m}{d} = \Theta(1)$.
Taking the zero temperature limit, i.e. $\beta \to \infty$, in the above equation, we finally obtain the ground-state energy $e_{\rm gs}$, a quantity commonly used in physics. Interestingly, we recognize the definition of the Rademacher complexity $\mathfrak{R}_m(\mF)$
\begin{align}
\begin{aligned}
   	e_{\rm gs} &\equiv \lim_{\beta \to \infty} \lim_{d \to \infty} \EE_{\vec{y}, \mat{X}} \left \langle  \frac{ 	\mH\(\{\vec{y},\mat{X}\}, \vec{w}\) }{d} \right\rangle_\beta &= \frac{\alpha}{2} \[ 1 - \EE_{\vec{y}, \mat{X}} \sup_{f_{\vec{w}}\in \mF}  \frac{1}{m} \sum_{\mu=1}^m y^{(\mu)} f_{\vec{w}}\(\vec{x}^{(\mu)}\)  \] \\
   	&=  \frac{\alpha}{2} \[ 1 -  \mathfrak{R}_m\(\mF\) \]\,,
   	\label{main:link_gs_rademacher}
\end{aligned}
\end{align}
where random labels $\vec{y}$ play the role of the Rademacher variable $\bepsilon$ in \eqref{main:def_rademacher}. The above equation shows a simple correspondence between the ground-state energy of the perceptron model and the Rademacher complexity of the corresponding hypothesis class, and shall bring insights from both the machine learning and statistical physics communities. Consequently, as we shall see, this connection means that the Rademacher complexity can be computed (rather than bounded) for many models using the replica method from statistical physics. As far as we are aware, this basic connection between the ground state energy and Rademacher complexity was not previously stated in literature. 

\subsection{An intuitive understanding on the Rademacher bounds on generalization}
At this point, the Rademacher complexity becomes a more familiar object to the physics-minded reader. However, could we understand more intuitively why the Rademacher complexity, or equivalently the ground-state energy, is involved in the generalization gap bound? Let us present an intuitive hand-waving explanation. Consider the fraction of mistakes performed by a classifier $f_{\vec{w}}$ on unknown samples, namely the generalization error $\epsilon_{\rm gen}(f_{\vec{w}})$, and on the training set, the training error $\epsilon_{\rm train}^m(f_{\vec{w}})$. The worst case scenario that could occur is trying to fit while there exists no underlying rule, meaning that labels are purely random uncorrelated from input. The estimator will purely overfit and its generalization error will remain constant to $1/2$ in any case. This leads to the following heuristic generalization bound:
\begin{align}
\begin{aligned}
   \epsilon_{\rm gen}(f_{\vec{w}}) - \epsilon_{\rm train}^m(f_{\vec{w}}) & \leq \epsilon^{\rm random~labels}_{\rm gen}(f_{\vec{w}}) - \epsilon^{{\rm random~labels}, m}_{\rm train}(f_{\vec{w}}) = \frac{1}{2} -  \epsilon^{{\rm random~labels}, m}_{\rm train}(f_{\vec{w}}) \\
   &= \frac{1}{2} \(1 - 2\epsilon^{{\rm random~labels}, m}_{\rm train}(f_{\vec{w}}) \) = \frac{1}{2} \hat{\mathfrak{R}}_m\(\mF\)\,.
\end{aligned}
\end{align}
Note that this heuristic reasoning does not give the {\it exact} Rademacher generalization bound. In fact, the actual stronger and uniform (over all possible $\vec{w}\in\bbR^{d}$) bound does not have a factor $1/2$, and surely cannot be fully captured by the simple above argument. Nevertheless, this argument reflects the crux of the Rademacher bound: it provides a very pessimistic bound by assuming the worst possible scenario: i.e. fitting data and trying to make predictions while the labels are random. Of course, in real data problems the rule is not random; it is then no surprise that the Rademacher bound is not tight (\cite{zhang2016understanding}). Indeed, real problems labels are \emph{not} randomly correlated with the inputs.

%%% Applications and figures %%%
\section{Consequences and bounds for simple models\label{sec:applications}}
In this section, we illustrate our previous arguments and the connection between the spin glass approach and the Rademacher complexity still for the case of Gaussian \iid input data matrix $\mat{X}$ in the high-dimensional limit when $m,d \to \infty$.

\subsection{Ground state energies of the perceptron}
\label{magic}
For a number of samples smaller than the Gardner capacity $\alpha_c$, it is by definition possible to fit all random labels $\vec{y}$. Accordingly, the number of misclassified examples is zero and the ground state energy $e_{\rm gs}=0$. This means that the Rademacher complexity is asymptotically equal to $1$ for $\alpha<\alpha_{c}$. However above the Gardner capacity $\alpha > \alpha_c$, the estimator $f_{\vec{w}}$ cannot perfectly fit the random labels and will misclassify some of them, equivalently $e_{\rm gs}>0$. From the arguments given in sec.~\ref{sec:iid}, we thus expect 
\begin{equation}
\begin{aligned}
	\mathfrak{R}_m \(\mF \) &= 1 ~~\text{for}~~ \alpha < \alpha_c\,,\\
	\mathfrak{R}_m \(\mF \) &\approx \Theta\(\sqrt{\frac{\alpha_c}{\alpha}} \) ~~\text{for}~~ \alpha \gg \alpha_c\,.
\end{aligned}
\end{equation}
This relation is already non-trivial, as it yields a link between the Gardner capacity and the Rademacher complexity. 
Using the replica method from spin glass analysis, and the mapping with ground state energies \eqref{main:link_gs_rademacher}, we shall now see how one can go beyond these simple arguments, and compute the actual precise asymptotic value of the Rademacher complexity.

\subsection{Computing the ground-state energy with the replica method}
Knowing that statistical physics literature focused mainly on the Gardner capacity, the connection between the ground-state energy and the Rademacher complexity suggests that it would be worth looking at these old results in a new light. In fact, the replica method allows for an {\it exact} computation of the Rademacher complexity for random input data in the large size limit. In the following, we handle computations by focusing on a simple generalization of the linear functions hypothesis class. Fix any activation function ${\varphi:\bbR \mapsto \{\pm 1\}}$, we define the following hypothesis class
\begin{align}
    \mF_\varphi \equiv \left\lbrace f_{\vec{w}}: 
\begin{cases}
\bbR^{d} \mapsto \{-1,1\}\\
\vec{x} \mapsto \varphi \( \frac{1}{\sqrt{d}} \vec{w}^\intercal \vec{x}\)
\end{cases}  , \vec{w} \in \bbR^{d} \right\rbrace \,. 
\label{main:glm_hypothesis_class}
\end{align}
Starting with the posterior distribution
\begin{align}
	\bbP ( \vec{w} | \vec{y}, \mat{X} ) = \frac{\bbP(\vec{y} | \vec{w}, \mat{X}) \bbP(\vec{w})  }{ \bbP ( \vec{y}, \mat{X} )  } = \frac{ e^{-\beta \mH( \{\vec{y}, \mat{X}\}, \vec{w})} P_w\( \vec{w} \)}{\mZ( \{\vec{y}, \mat{X}\}, \alpha, \beta)} \,,
\end{align}
we introduced the partition function associated to the Hamiltonian eq.~\eqref{main:hamiltonian} at inverse temperature $\beta$
\begin{align}
	\mZ( \{\vec{y}, \mat{X}\}, \alpha, \beta) = \int_{\bbR^d} \d \vec{w} e^{-\beta \mH( \{\vec{y}, \mat{X}\}, \vec{w})} P_w\( \vec{w} \)\,.
	\label{main:partition_function}
\end{align}
In the large size limit $d \to \infty$, the posterior distribution becomes highly peaked in particular regions of parameters. In physics we are interested in these dominant regions and focus on the free energy at inverse temperature $\beta$ defined as
\begin{align}
    \Phi_{\vec{y}, \mat{X}} (\{\vec{y}, \mat{X}\}, \alpha, \beta) \equiv - \lim_{d \to \infty} \frac{1}{d \beta} \log \mZ( \{\vec{y}, \mat{X}\}, \alpha, \beta) \,.
    \label{main:free_energy_not_avg}	
\end{align}
However, as we are interested in computing quantities in the \emph{typical case}, we want to average over all potential training sets $\{\vec{y}, \mat{X}\}$ and compute instead the averaged free energy
\begin{align}
\Phi(\alpha, \beta) \equiv \EE_{\vec{y}, \mat{X}} \[ \Phi_{\vec{y}, \mat{X}} \( \{\vec{y}, \mat{X}\}, \alpha, \beta\) \].
\end{align}
Computing directly this average rigorously is difficult, hence we will carry out the computation using the so-called \emph{replica method}, starting by writing the \emph{replica trick}
\begin{align}
    - \frac{1}{d \beta} \EE_{\vec{y}, \mat{X}} \[ \log \mZ( \{\vec{y}, \mat{X}\}, \alpha, \beta) \] = - \frac{1}{d \beta}  \lim_{r \to 0}  \frac{\partial \log \EE_{\vec{y}, \mat{X}} \[ \mZ(\{\vec{y}, \mat{X}\}, \alpha, \beta)^r\] }{\partial r} \,,
    \label{main:replica_trick}	
\end{align}
which replaces the expectation of $\EE\[\log \mZ\]$ by the moments of $\EE\[\mZ^r\]$, which are easier to compute. 
For the non-familiar reader, it is instructive to remark that we introduced $\mZ^r$ the partition function of $r \in \bbN$ non-interacting copies, also called \emph{replicas}, of the initial system with partition function $\mZ$. Assuming there exists an analytical continuation $r\in \bbR$ and that we can revert both limits, we can finally take the limit $r \rightarrow 0$
\begin{align}
    \Phi(\alpha, \beta) = \lim_{r \to 0} \[ \lim_{d \to \infty}  - \frac{1}{d \beta}  \frac{\partial  \log \EE_{\vec{y}, \mat{X}} \[ \mZ(\{\vec{y}, \mat{X}\}, \alpha, \beta)^r\] }{\partial r} \] \,.
    \label{main:free_energy_trick}
\end{align}

We give some details on the replica computation in Appendix~\ref{appendix:replicas_iid}, and we also refer the reader to the relevant literature in physics (\cite{Mezard1986,Hertz1993,engel2001statistical,mezard2009information,Zdeborova2016}) and in mathematics (\cite{talagrand2003spin,talagrand2006parisi,bolthausen2007spin,panchenko2004bounds,panchenko2018free}). 
The computation of the high-dimensional posterior and corresponding free energy $\Phi$ \eqref{main:free_energy_trick} reduces to an optimization problem over two symmetric matrices $\mat{Q}, \hat{\mat{Q}} \in \bbR^{r\times r}$ which describe the correlations (induced by the average) between the aforementioned fictive \emph{replicas}. In particular the off-diagonal terms of the matrix $\mat{Q}=\(\frac{1}{d} \vec{w}^a \cdot \vec{w}^b \)_{a,b=1}^r$ measure the \emph{overlaps} between the different replicas, while the diagonal term is fixed to $\EE\[\frac{1}{d}\|\vec{w}\|_2^2\]$. We can therefore derive the free energies corresponding to
a hierarchy of approximate ansatz on these matrices $\mat{Q}$ and $\hat{\mat{Q}}$, named \emph{replica symmetric} ({\rm{RS}}), \emph{one-step replica-symmetry breaking} ({\rm{1RSB}}), \emph{two-step replica-symmetry breaking} ({\rm{2RSB}})\ldots 
The different ansatz describe different solution space structures and we refer the interested reader to Appendix \ref{appendix:replica:rs} and \ref{appendix:replica:rs} for more details. 
While in some problems the {\rm{RS}} or the {\rm{1RSB}} ansatz are sufficient, in others only the infinite step solution (full-{\rm{RSB}})  gives the exact ansatz (\cite{Mezard1989,talagrand2003spin,talagrand2006parisi}), although the 1RSB approach is usually an accurate approximation. 

Computing the ground state energy consists in taking the zero temperature limit $\beta \to \infty$ above the capacity $\alpha>\alpha_c$ in the replica free energy $\Phi(\alpha,\beta) = e(\alpha,\beta) - \beta^{-1} s(\alpha,\beta)$; where $e, s$ denote respectively the energy and entropy contributions. The simplest form of the replica computation is known as \emph{Replica Symmetric} ({\rm RS}) and the next simplest is \emph{one-step Replica-Symmetry Breaking} ({\rm 1RSB}) which plugged in eq.~\eqref{main:free_energy_trick} leads to expressions \cite{Majer1993a, Erichsen1992, Whyte1996}
\begin{align}
\begin{aligned}
\label{main:free_energy_iid}
\Phi^{(\rm rs)}_{\rm iid} (\alpha, \beta) &= -\frac{1}{\beta} \underset{ q_0, \hat{q}_0}{\textbf{extr}}     \left\{\frac{1}{2}\( q_0\hat{q}_0 -1 \)+  \Psi_{\rm w}^{(\rm rs)}(\hat{q}_0)   +\alpha  \Psi_{\rm out}^{(\rm rs)}(q_0, \beta)    \right\} \,, \\ 
\Phi^{(\rm 1rsb)}_{\rm iid}(\alpha, \beta) &=  - \frac{1}{\beta}\underset{ q_0, q_1, \hat{q}_0, \hat{q}_1, x_0}{\textbf{extr}} \left\{  \frac{1}{2} \left(  q_1\hat{q}_1 - 1 \right) + \frac{x_0}{2} \left(q_0\hat{q}_0 - q_1\hat{q}_1 \right)   \right. \\
 &	\left. \hspace{3cm} + \Psi_{\rm w}^{(\rm 1rsb)}(\hat{q}_0, \hat{q}_1)   +\alpha \Psi_{\rm out}^{(\rm 1rsb)}(q_0, q_1,\beta)    \right\}  \,,
\end{aligned}
\end{align}
where $q_0, q_1$ denote the overlap order parameters and the auxiliary functions
\begin{align}
\begin{aligned}
    \Psi_{w\rm }^{\rm (rs)}(\hat{q}_0) &\equiv \EE_{\xi_0}  \log   \EE_{w} \[ \exp \(             {\frac{(1- \hat{q}_0  )}{2}  w^2}+ \xi_0 \sqrt{\hat{q}_0} w \)  \]  \,, \\
	\Psi_{\rm out}^{\rm (rs)}(q_0, \beta) &\equiv  \EE_y \EE_{\xi_0}  \log  \EE_{z} \[\mI \(y \big | \sqrt{Q - q_0} z + \sqrt{q_0}\xi_0, \beta \)  \]\,, \\
	\Psi_{\rm w}^{(\rm 1rsb)}(\hat{q}_0,\hat{q}_1) &\equiv \frac{1}{x_0}
                \EE_{\xi_0} \log\left( \EE_{\xi_1} \EE_w \[\exp\left(
                  \frac{(1- \hat{q}_1  )}{2} w^2 +
                  \left(\sqrt{\hat{q}_0}\xi_0+\sqrt{\hat{q}_1-\hat{q}_0}\xi_1
                  \right)w \right)  \]^{x_0}\right)  \,,\\
    \Psi_{\rm out}^{(\rm 1rsb)}(q_0, q_1,\beta) &\equiv \frac{1}{x_0}
                \EE_y \EE_{\xi_0}  \log\left(  \EE_{\xi_1} \EE_{z} \[\mI( y \big | \sqrt{q_0} \xi_0 +
                \sqrt{q_1-q_0} \xi_1  + \sqrt{1-q_1} z,\beta) \]^{x_0} \right)\,,   
\end{aligned}
\end{align} 
where $\xi_0, \xi_1$ denote two \iid normal random variables, and $y\sim P_y(.)$ the distribution of the random labels.
We introduced a temperature-dependant constraint function $\mI(y|z)=e^{- \beta V(y|z)}$ where the generic cost function $V$ reads in our case $V(y|z) = \id\[ y  \ne \varphi (z) \]$. Above expressions are valid for any generic weight distribution $P_w(.)$ and non-linearity $\varphi$. The detailed computation can be found in Appendix~\ref{appendix:replicas_iid}, in particular eq.~\eqref{appendix:free_energy_rs} and eq.~\eqref{appendix:free_energy_1rsb}. Then the general method to find the ground state energy is to take the zero temperature limit
\begin{align}
\label{main:def_gs}
    e_{\rm gs, iid}(\alpha) \equiv \lim_{\beta \to \infty} \Phi_{\rm iid} (\alpha, \beta) \,,
\end{align}
while handling carefully the scaling of the optimized order parameters in this limit.
\begin{figure}[t] 
\centering
    \includegraphics[scale=0.6]{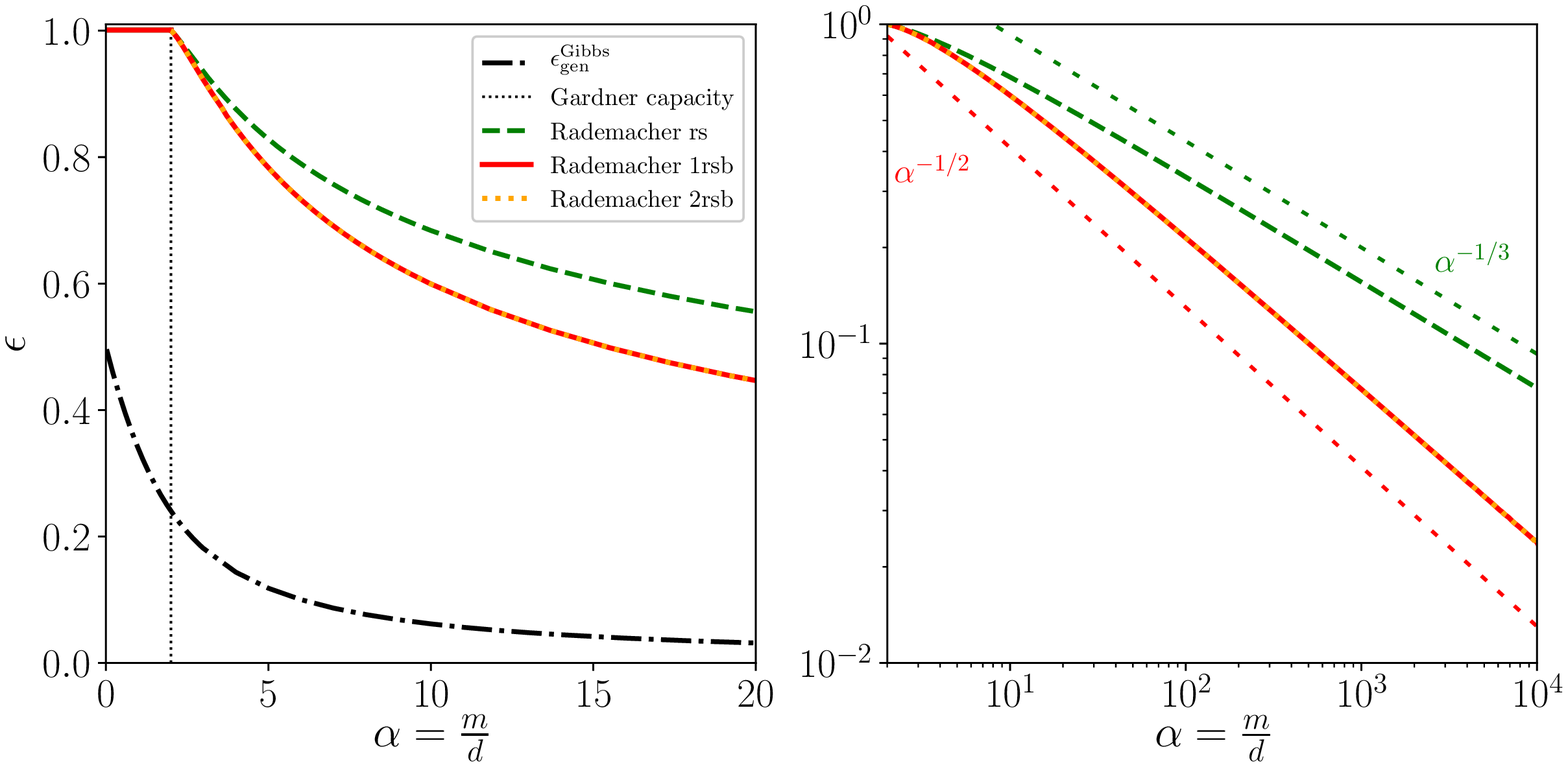}
    \caption{
    Explicit Rademacher complexity for the spherical perceptron ($\alpha_c=2$). For $\alpha<\alpha_c$ the problem is satisfiable so the number of error is zero and the Rademacher complexity is constant to unity. For $\alpha> \alpha_c$, the problem becomes unsatisfiable and $e_{\rm gs}>0$. In the case of the spherical perceptron, RS (dashed green) and 1RSB (red) ansatz provide really different results that scale respectively with $\alpha^{-1/3}$ and $\alpha^{-1/2}$ (scaling are represented with colored dashed lines). Performing {\rm{2RSB}} (dashed orange, see Appendix \ref{appendix:replicas_ground_state_spherical}) does not change the scaling and the difference with respect to {\rm{1RSB}} is visually imperceptible.
    The black dotted-dashed curve is the generalization error in the teacher-student scenario \cite{Barbier2017b}. Note the large gap between the worst case Rademacher bound and the actual teacher-student generalization error.}
    \label{fig:rademacher_spherical}
\end{figure}
\paragraph{Spherical perceptron}
The most commonly studied model (\cite{gardner1988optimal,Gardner1988a,Gardner1989,gardner1988optimal}) with continuous weights is the spherical model with $\vec{w}\in\bbR^{d}$ such that $\|\vec{w}\|_2^2 = d$. The spherical constraint allows to have a well-defined model which excludes diverging or vanishing weights. In this case, the Gardner capacity is rigorously known to be equal to $\alpha_c=2$ (\cite{cover1965geometrical}). 

We computed the {\rm{RS}}, {\rm{1RSB}} and {\rm{2RSB}} free energies (\cite{Majer1993a, Erichsen1992, Whyte1996}, see details in Appendix \ref{appendix:replicas_ground_state_spherical}.). Taking the zero temperature limit $\beta \to \infty$ with $q_0 \to 1$ and $\chi = \beta (Q-q_0)$ finite in the RS ansatz and $q_1 \to 1, x_0\to 0$  keeping $\chi \equiv \beta (Q-q_1)$ and $\Omega_0 \equiv \frac{x_0 \beta}{\chi}$ finite in the {\rm{1RSB}} case, leads to the following expressions of the ground states energies:
\begin{align}
\label{main:ground_states_spherical}
    e_{\rm gs, iid}^{\rm (rs)} &=  \extr_{\chi} \left\{ -\frac{1}{2 \chi} + \alpha  \EE_{y, \xi_0} \min_{z} \[ V(y|z) + \frac{\(z - \xi_0\)^2}{2 \chi}  \]   \right\} \spacecase
    e_{\rm gs, iid}^{\rm (1rsb)} &= \extr_{\chi,\Omega_0,q_0} \left\{  \frac{1}{2\Omega_0 \chi} \log\( 1 + \Omega_0 (1-q_0) \) + \frac{q_0}{2\chi \( 1 + \Omega_0 (1-q_0) \) }   \right. \\
& \left. \hspace{4cm}  + \frac{\alpha}{\chi \Omega_0}  \EE_{\xi_0}  \log \EE_{\xi_1} e^{-\Omega_0 \chi \min_z \[V(y | z) + \frac{1}{2 \chi } \(z - \sqrt{q_0} \xi_0 -
                \sqrt{1-q_0} \xi_1 \)^2 \]}   \right\}\,, \nonumber
\end{align}

where the cost function $V(y|z) = \id\[ y  \ne \varphi (z) \]$. The details of the  derivation via the replica method and the expression for the 2RSB ansatz are given in Appendix \ref{appendix:replicas_ground_state_spherical}. The results for Rademacher variable $y$ and with $\varphi(z) = \sign(z)$ are depicted in Fig.~\ref{fig:rademacher_spherical}. 

Interestingly, the bounds on the Rademacher complexity also induce consequences on spin glass physics. 
Indeed as the large $\alpha$ scaling of the ground state energy is not generally prescribed,
the fact that the Rademacher complexity scales as $\alpha^{-1/2}$ for large values of $\alpha$ --- namely there exists a constant $\mC$ such that $\mathfrak{R}_m \( \mF \) \underset{\alpha \to \infty}{\approx} \frac{\cal C}{\sqrt{\alpha}}$ --- implies that the ground state energy behaves for large $\alpha$ as
\begin{equation}
e_{\rm gs} (\alpha) = \frac{\alpha}{2} \(1-\mathfrak{R}_m \( \mF \)\) \underset{\alpha \to \infty}{ \longrightarrow} \frac{\alpha}2 \left(1-\frac{\cal C}{\sqrt{\alpha}}\right)\,.
\label{scaling-rsb-from-VC}
\end{equation}

We first notice that the replica symmetric ({\rm{RS}}) solution complexity fails to deliver the correct scaling as sketched in Fig.~\ref{fig:rademacher_spherical}, so the scaling in eq.~(\ref{scaling-rsb-from-VC}) must not be entirely trivial. On the other hand, the {\rm{1RSB}} and {\rm{2RSB}} solutions we used (which are expected to be numerically very close to the harder to evaluate full-{\rm{RSB}} one), seems to yield the correct scaling (see Fig.~\ref{fig:rademacher_spherical}). 
It is rather striking that the statistical learning connection allows to predict, through eq.~(\ref{scaling-rsb-from-VC}), the scaling of the energy in the large $\alpha$ regime, that is only satisfied with replica-symmetry breaking ansatz. This yields an open question for replica theory: in practice, can one compute exactly the value of the constant $\cal C$? Given the full-RSB solution is notoriously hard to evaluate, this might be an issue worth investigating in mathematical physics.

\paragraph{Binary perceptron}
Another common choice for the weights distribution is the binary prior\\ ${P_w(w)= \delta(w-1) + \delta(w+1)}$ studied e.g. in \cite{Krauth1989}. In this case, the Gardner capacity is predicted to be $\alpha_c \approx 0.83\ldots$, a prediction which, remarkably, is still not entirely rigorously proven, but see \cite{Sun2018,Aubin2019}.

\begin{figure}[t]
\centering
    \includegraphics[scale=0.6]{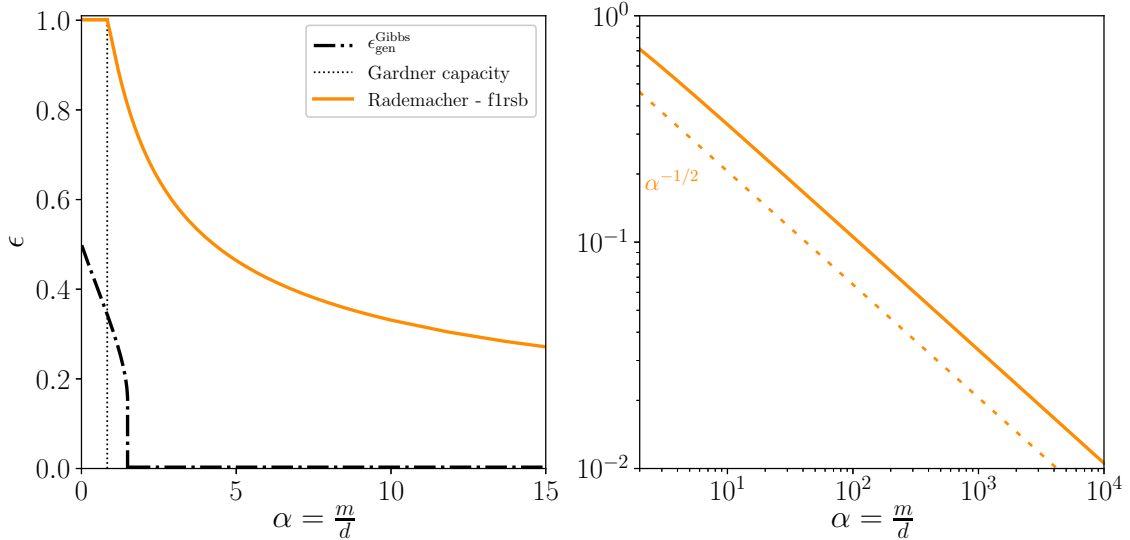}
    \caption{
    Explicit Rademacher complexity for (\textbf{left}) the binary perceptron ($\alpha_c=0.83\ldots$). The replica solution (orange) leads again (\textbf{right}) to a $\alpha^{-1/2}$ scaling (dashed orange) of the Rademacher complexity at large $\alpha$. The dotted-dashed black curve is the generalization error in the teacher-student scenario. Note the gap between the worst case bound (Rademacher) and the teacher-student generalization error.
    }
    \label{fig:rademacher_binary}
\end{figure}

To see this, we use eq.~\eqref{main:free_energy_iid}. In the binary perceptron, the landscape of the model is said to be \emph{frozen} 1RSB (f1RSB), i.e. clustered in point-like dominant solutions, and the {\rm{RS}} and {\rm{1RSB}} free energies are the same (even though their entropies are different) $\Phi(\alpha,\beta) = e(\alpha,\beta) - \beta^{-1} s(\alpha,\beta)$. In this case computing the ground state can be tackled via finding the effective temperature $\beta^\star$ such that the $s(\alpha, \beta^\star) = 0$, that can be plugged back to find the ground state energy $e_{\rm gs}(\alpha) =\Phi(\alpha,\beta^\star)$. Again, we note that even though the {\rm{1RSB}} ansatz is unstable and should be replaced by a more complex (and ultimately full-{\rm{RSB}}) solution, it already gives the good scaling $\mathfrak{R}_m(\mF) \sim \alpha^{-1/2}$, and satisfies the scaling eq.~(\ref{scaling-rsb-from-VC}) for large $\alpha$, as in the case of the spherical model, see Fig.~\ref{fig:rademacher_binary}.

\subsection{Teacher-student scenario versus worst case Rademacher}

The Rademacher bounds are really interesting as they depend only on the data distribution, and are valid for \emph{any rule} used to generate the labels, no matter how complicated. In this sense, it is a worst-case scenario on the rule that prescribes labels to data. A different approach, again pioneered in statistical physics \cite{Gardner1989}, is to focus on the behavior for a given rule, called the \emph{teacher} rule. Given the Rademacher bounds tackle the worst case with respect to that rule, it is interesting to consider the generalization error one actually gets for the \emph{best case}, i.e. fitting the labels according to the same teacher rule..  This is the  so-called \emph{teacher-student} approach.
%, where labels are generated by feeding \iid random samples to a neural network architecture (the teacher) and are then presented to another neural network (the student) that is trained using these data. 
In the wake of the need to understand the effectiveness of neural networks, and the limitations of the classical approaches, it is of interest to revisit the results that have emerged thanks to the physics perspective. 

We shall thus assume that the {\it actual labels} are given by the rule
\begin{equation}
    y = {\rm{sign}}{\left( \frac{1}{\sqrt{d}} \vec{w}^{\star \intercal} \vec{x}\right)}\,,
\end{equation}
with $\vec{w}^\star$, the \emph{teacher weights} that can be taken as Rademacher $\pm1$ variables, or Gaussian ones. Now that labels are generated by feeding \iid random samples to a neural network architecture (the teacher) and are then presented to another neural network (the student) that is trained using this data, it is interesting to compare the worst case Rademacher bound with the actual generalization error of this student on such synthetic data.

We now  consider the error of a \emph{typical} solution $\vec w$ from the posterior distribution (this is often called the Gibbs rule) for the student. Given the rule is outputting $\pm 1$ variables, this yields
\begin{align}
    \epsilon_{\rm gen}^{\rm Gibbs} = 1 - \EE_{\vec{x},\vec{w}^\star} \[ \langle  f_{\vec{w}^\star}(\vec{x}) \times f_{\vec{w}}(\vec{x}) \rangle \] = 1- q^\star
\end{align}
where $q^\star=\EE_{\vec{x},\vec{w}^\star} \[ \langle  f_{\vec{w}^\star}(\vec{x}) \times f_{\vec{w}}(\vec{x}) \rangle \] $. Computing the replica symmetric overlap $q^\star$ can be done within the statistical mechanics approach (\cite{Seung1992,Watkin1993,opper1995statistical,engel2001statistical}) and can be rigourously done as well (\cite{Barbier2017b}). Notice that this error is equal to the Bayes optimal error for the quadratic loss (see as well \cite{Barbier2017b}).

The two \emph{optimistic} (teacher-student) and \emph{pessimistic} (Rademacher) errors can be seen in Fig.~\ref{fig:rademacher_spherical} for spherical and in Fig.~\ref{fig:rademacher_binary} for binary weights. In this case, since a perfect fit is always possible, the training error is zero and the Rademacher complexity is itself the bound on the generalization error. These two figures show how different the worst and teacher-student case can be in practice, and demonstrate that one should perhaps not be surprised by the fact that the empirical Rademacher complexity does not always give the correct answer \cite{zhang2016understanding}, as after all it deals only with worst case scenarios.
\subsection{Committee machine with Gaussian weights}
Given the large gap between the Rademacher bound and the teacher-student setting, we can ask wheather we can find a case where the Rademacher bound is void in the sense that the Rademacher complexity is $1$ yet generalization is good for the teacher-student setting? This can be done by moving to two-layer networks. Consider a simple version of this function class, namely the committee machine \cite{engel2001statistical}. It is a two-layer network where the second layer has been fixed, such that only weights of the first layer $\mat{W}= \{\vec{w}_1, \cdots,\vec{w}_K \}\in\bbR^{d \times K}$ are learnt. The function class for a committee machine with $K$ hidden units is defined by
\begin{align}
    \mF_{\rm com} \equiv \left\lbrace f_{\mat{W}}: 
\begin{cases}
\bbR^{d} \mapsto \{-1,1\}\\
\vec{x} \mapsto {\rm{sign}}{\left( \sum_{k=1}^K   {\rm{sign}}{\left( \frac{1}{\sqrt{d}} \vec{w}_k^\intercal \vec{x}\right)}\right)}
\end{cases}   \mat{W}\in\bbR^{d \times K} \right\rbrace \,. 
\label{eq:comm}
\end{align}
Instead of computing the Rademacher complexity with the replica method, it is sufficient for the purpose of this section to understand its rough behavior. As discussed in sec.~\ref{magic}, this requires knowing the Gardner capacity. A generic bound by \cite{Mitchison1989} states that it is upper bounded by $\Theta(K \log(K))$. Additionally, the Gardner capacity has been computed by the replica method in \cite{Monasson_1995,urbanczik1997storage,xiong1998storage} who obtained  that $\alpha_c = \Theta(K \sqrt{\log(K)})$. We thus expect that 
\begin{equation}
\begin{aligned}
	\mathfrak{R}_m \(\mF_{\rm com} \) &= 1 ~~\text{for}~~ \alpha < \Theta\(K \sqrt{\log(K)})\) \,,\\
	\mathfrak{R}_m \(\mF_{\rm com} \) &\approx \Theta  \(\sqrt{\frac{K\sqrt{\log{K}}}{\alpha}}\) ~~\text{for}~~ \alpha \gg \Theta \(K \sqrt{\log{K}}\)\,.
\end{aligned}
\end{equation}

\begin{figure}[t]
\centering
    \includegraphics[scale=0.6]{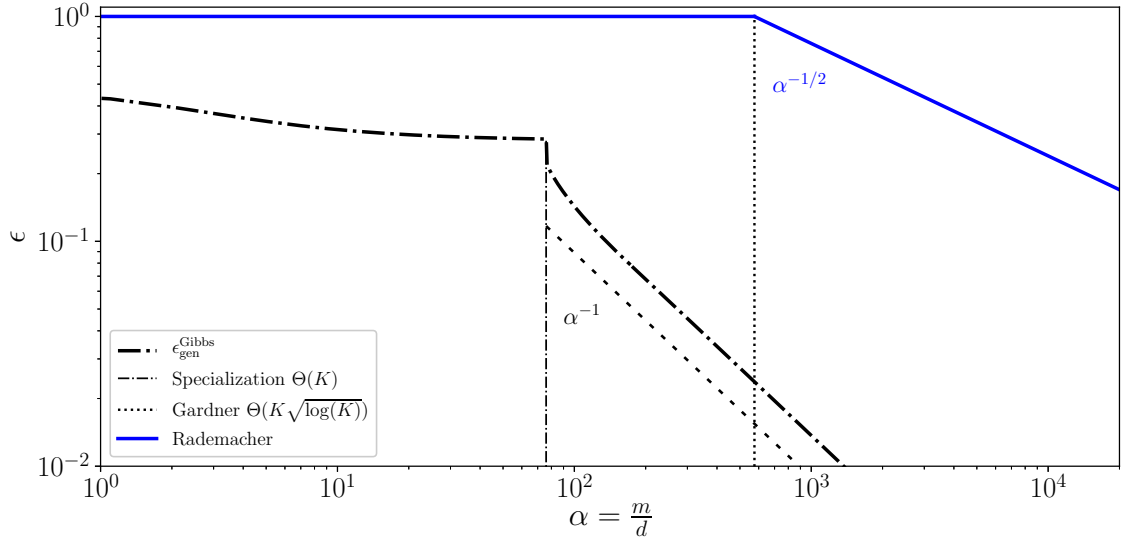}
    \caption{Illustration of the scaling of the Rademacher complexity (blue) for the fully connected committee machine, drawn together with the exact generalization error in the teacher-student scenario (dotted-dashed black), scaling as $\alpha^{-1}$ at large $\alpha$. Notice the large gap between the worst case bound (Rademacher) and the teacher-student result.}
    \label{fig:rademacher_committee}
\end{figure}

To compare with the teacher-student case, when the labels are  produced by a teacher committee machine as
\begin{equation}
      y = {\rm{sign}}{\left( \sum_{k=1}^K   {\rm{sign}}{\left( \frac{1}{\sqrt{d}} \vec{w}^{\star \intercal}_k \vec{x}\right)}\right)}\,,
\end{equation}
the error of the Gibbs algorithm reads
\begin{align}
    \epsilon_{\rm gen}^{\rm Gibbs} = 1 - \EE_{\vec{x},\vec{w}^\star} \[ \langle  f_{\vec{w}^\star}(\vec{x}) \times f_{\vec{w}}(\vec{x}) \rangle \] = 1- q^\star
\end{align}
where, again $q^\star=\EE_{\vec{x},\vec{w}^\star} \[ \langle  f_{\vec{w}^\star}(\vec{x}) \times f_{\vec{w}}(\vec{x}) \rangle \] $, has been computed in a series of papers in statistical physics \cite{Hertz1993,schwarze1993learning}, and using the Guerra interpolation method in \cite{Aubin2018}. Interestingly, in this case, one can get an error that decays as $1/\alpha$ as soon as $\alpha \gg \Theta(K)$. One thus observes a huge gap between the Rademacher bound that scales as $ \mathfrak{R}_m\(\mF_{\rm com}\) =\Theta\(\sqrt{K\sqrt{\log(K)}/\alpha}\)$ and the actual generalization error $\epsilon_g = \Theta(K/\alpha)$ for large sample size. This large gap further illustrates the considerable difference in behavior one can get between the worst case and teacher-student case analysis, see Fig.~\ref{fig:rademacher_committee}.

\subsection{Extension to rotationally invariant matrices}
\label{sec:rot_inv_mat}
The previous computation for \iid data matrix \mat{X} can be generalized to rotationally invariant (\rm{RI}) random matrices $\mat{X} = \mat{U}\mat{S}\mat{V}$ with rotation matrices $\mat{U} \in \mat{O}\(d\)$, $\mat{V} \in \mat{O}\(m\)$ independently sampled from the Haar measure, and $\mat{S} \in \bbR^{d \times m}$ a diagonal matrix of singular values. Computation for this kind of matrices can be handled again using the replica method (\cite{Kabashima2008,barbier2018mutual,Gabrie2018}) and leads to {\rm{RS}} and {\rm{1RSB}} free energies
\begin{align}
\begin{aligned}
\Phi^{\rm (rs)}_{\rm RI}(\alpha, \beta) &= - \dfrac{1}{\beta} \extr_{\chi_w, \chi_u, q_w, q_u} \left\lbrace  \mathcal{A}_0^{\rm (rs)} (\chi_w, \chi_u, q_w, q_u)  \right.   \\
    & \left. \hspace{3cm} + \mathcal{A}_w^{\rm (rs)} (\chi_w, q_w) + \alpha \mathcal{A}_u^{\rm (rs)} (\chi_u, q_u, \beta)\right\rbrace\,, \\
\Phi^{\rm(1rsb)}_{\rm RI} (\alpha, \beta) &=- \dfrac{1}{\beta} \extr_{\chi_w, \chi_u, v_w, v_u, q_w, q_u, x} \left\lbrace \mathcal{A}_0^{\rm (1rsb)} (\chi_w, \chi_u, v_w, v_u, q_w, q_u, x) \right.  \\ 
    & \left. \hspace{3cm} + \mathcal{A}_w^{\rm (1rsb)}(\chi_w, v_w, q_w, x)  + \alpha \mathcal{A}_u^{\rm (1rsb)}(\chi_u, v_u, q_u, x, \beta)\right\rbrace\,,
    \label{main:free_energy_RI}
\end{aligned}
\end{align}
where each term is properly defined in Appendix \ref{appendix:replicas_rot_inv_mat}. Note that taking $\mat{X}$ a random Gaussian \iid matrix, the eigenvalue density $\rho(\lambda)$ follows the Marchenko-Pastur distribution and~\eqref{main:free_energy_RI} matches free energies eq.~\eqref{appendix:free_energy_rs}, \eqref{appendix:free_energy_1rsb}, and ground states energies eq.~\eqref{main:ground_states_spherical} in the spherical case.  The ground state energy (and therefore the Rademacher complexity) can be again computed as in the \iid case, taking the zero temperature limit $\beta \to \infty$
\begin{align}
    e_{\rm gs, RI}(\alpha) = \lim_{\beta \to \infty}   \Phi_{\rm RI}(\alpha, \beta)\,,
\end{align}
keeping in particular $\beta \chi_w$ and $x \beta$ finite in the limits $\beta \to \infty, x\to 0, \chi_w \to 0$.

%%% Discussion and consequences %%%
\section{Conclusion \label{sec:discussion}}
In this paper, we discussed the deep connection between the Rademacher complexity and some of the classical quantities studied in the statistical physics literature on neural networks, namely the Gardner capacity, the ground-state energy of the random perceptron model, and the generalization error in the teacher-student model.  We believe it is rather interesting to draw the link with approaches inspired by statistical physics, and compare its findings with the worst-case results. In the wake of the need to understand the effectiveness of neural
networks and also the limitations of the classical approaches, it is of interest to revisit the results that have emerged thanks to the physics perspective. This direction is currently experiencing a strong revival, see e.g. \cite{Chaudhari2016,martin2017rethinking,Advani2017a,Baity-Jesi2018}. The connection discussed in the paper opens the way to a unified presentation of these often contrasted approaches, and we hope this paper will help bridging the gap between researchers in traditional statistics and in statistical physics.  There are many possible follow-ups, the more natural one being the computation of Rademacher complexities from statistical physics methods for more complicated and realistic models of data, starting for instance with correlated matrices discussed in section \ref{sec:rot_inv_mat}.

\section{Acknowledgements}
This work is supported by the ERC under the European Union’s Horizon
2020 Research and Innovation Program 714608-SMiLe, as well as by the
French Agence Nationale de la Recherche under grant
ANR-17-CE23-0023-01 PAIL and from the Chaire CFM-ENS.
We thank Henry Pfister for insightful and clarifying discussions that inspired partly this work. We would also like to thank the Kavli Institute for Theoretical Physics (KITP) for welcoming us during part of this research, with the support of the National Science Foundation under Grant No. NSF PHY-1748958.

%\clearpage
\bibliographystyle{unsrt}
\bibliography{rademacher_bib.bib}

\newpage
\appendix
\appendix
\section*{Appendix}
\label{appendix}

\section{Rademacher scaling for perceptron}
\label{appendix:rademacher_linear_models}
 
\begin{proof}
\subsubsection*{Upper bound}
For a linear classifier with binary outputs such as the perceptron, the VC dimension is easy to compute and $d_{\rm VC} = d$. Hence we know from Massart theorem's \cite{massart2000some} that
\begin{align*}
	\mathfrak{R}_m(\mF_{\rm sign}) \leq \Theta \(\sqrt{\frac{d_{\textrm{VC}}(\mF_{\rm sign})}{m}}\) = \Theta\(\sqrt{\frac{d}{m}}\) = \Theta\(\alpha^{-1/2}\)  \,.
\end{align*}

\subsubsection*{Lower bound}
Let us consider the following estimator (known as the Hebb's rule \cite{hebb1962organization}): $\displaystyle \vec{w}^\star = \frac{1}{\sqrt{d}} \sum_{\nu=1}^m y^{(\nu)} \vec{x}^{(\nu)}$. Hence for a given sample $\vec{x}^{(\mu)}$ the above estimator outputs
$$f_{\vec{w}^\star} \(\vec{x}^{(\mu)}\) = \sign\(  \frac{1}{\sqrt{d}} \vec{w}^{\star \intercal} \vec{x}^{(\mu)} \) = \sign \( \(\displaystyle  \frac{1}{d} \sum_{\nu=1}^m y^{(\nu)} \vec{x}^{(\nu)} \)^\intercal \vec{x}^{(\mu)} \)\,.$$ 
Injecting its expression in the definition the Rademacher complexity eq.~\eqref{main:def_rademacher} one obtains:
\begin{align*}
	&\mathfrak{R}_m(\mF_{\rm sign}) \equiv  \EE_{\vec{y}, \mat{X}} \[ \sup_{\vec{w}} \frac{1}{m} \sum_{\mu =1}^m  y^{(\mu)}  f_{\vec{w}} \(\vec{x}^{(\mu)}\) \] \geq   \EE_{\vec{y}, \mat{X}} \[ \frac{1}{m} \sum_{\mu=1}^m y^{(\mu)} f_{\vec{w}^\star}\(\vec{x}^{(\mu)} \) \] \\
	&=   \EE_{\vec{y}, \mat{X}} \[ \frac{1}{m} \sum_{\mu=1}^m \sign \( y^{(\mu)}  \frac{1}{d} \(\sum_{\nu=1}^m y^{(\nu)} \vec{x}^{(\nu)} \)^\intercal \vec{x}^{(\mu)}\) \] \\
	&=  \EE_{\vec{y}, \mat{X}} \[ \frac{1}{m} \sum_{\mu=1}^m \sign \(1 + \frac{1}{d}\sum_{\nu \ne \mu}^m y^{(\mu)}  y^{(\nu)} \vec{x}^{(\nu)\intercal} \vec{x}^{(\mu)} \)\] \,.
\end{align*}
As $\vec{x}^{(\mu)} \sim \mN\(\vec{0},\mat{I}_d\)$ and $y^{(\mu)} \sim \mathcal{U}(\pm 1)$, $\vec{z}^{(\mu)} \equiv y^{(\mu)} \vec{x}^{(\mu)} \sim \mN\(\vec{0},\mat{I}_d\)$.
Hence let us define the Gaussian random variable
\begin{align*}
    \theta_\mu \equiv \frac{1}{d}\sum_{\nu \ne \mu}^m y^{(\mu)}  y^{(\nu)} \vec{x}^{(\nu)\intercal} \vec{x}^{(\mu)} = \frac{1}{d}\sum_{\nu \ne \mu}^m \vec{z}^{(\nu)\intercal} \vec{z}^{(\mu)}\,,
\end{align*}
and compute its two first moments
\begin{align*}
	\EE \[\theta_{\mu}\] &= \EE_{\vec{z}}\[ \frac{1}{d}\sum_{\nu \ne \mu}^m \vec{z}^{(\nu)\intercal} \vec{z}^{(\mu)} \] = \EE_{\vec{z}}\[ \frac{1}{d}\sum_{\nu \ne \mu}^m  \sum_{i=1}^d z_i^{(\nu)\EE_{\vec{y}, \mat{X}}} z_i^{(\mu)} \] = 0 \,,\spacecase
	\EE \[\theta_{\mu}^2\] &= \EE\[ \frac{1}{d^2} \(\sum_{\nu \ne \mu}^m \vec{z}^{(\nu)\intercal} \vec{z}^{(\mu)}\)^2 \] = \frac{(m-1)}{d} \underset{m \to \infty}{\longrightarrow}  \alpha \,.
\end{align*}
Hence because of the central limit theorem, in the high-dimensional limit $\theta_\mu \sim \mN(0, \alpha)$. Finally
\begin{align*}
	\mathfrak{R}_m(\mF_{\rm sign}) &\geq  \EE_{\btheta} \[ \frac{1}{m} \sum_{\mu=1}^m \sign\(1 + \theta_\mu \)\] =\EE_{\theta} \[ \sign\( 1 + \theta\)  \] \\
	&=   \bbP\[ \theta \geq -1 \] -  \bbP\[\theta \leq -1 \] = 2 \bbP\[ \theta \geq -1 \] - 1 \,.
\end{align*}
Noting that 
\begin{align*}
	\bbP\[ \theta \geq -1 \] &= \int_{-\frac{1}{\sqrt{\alpha}}}^\infty \Diff_\theta = \frac{1}{2} {\rm erfc}\(- \frac{1}{\sqrt{2\alpha}} \) \underset{\alpha \to \infty}{\simeq} \frac{1}{2} - \frac{1}{\sqrt{2\pi \alpha}}\,,
\end{align*}
we obtain a lower bound for the Rademacher complexity
\begin{align*}
	\mathfrak{R}_m\(\mF_{\rm sign}\) \geq \sqrt{\frac{2}{\pi}} \frac{1}{\sqrt{\alpha}} = \Theta\(\frac{1}{\sqrt{\alpha}}\)\,.
\end{align*}
\end{proof}

%%%%%%%%%%%%%%%%%%%%%%%%%%%%%%%%%%%%%%%%%%%%%%%%%%%%%%%%%%%%%%%%%%%%%%%%%%%%%%%%%%%%%%%%%%%%%%%%
\section{Replica computation of the ground state energy for perceptrons}
\label{appendix:replica}

\subsection{Gaussian \iid matrix}
\label{appendix:replicas_iid}
In this section, we present the replica computation of \emph{Generalized Linear Models} (GLM) corresponding to the hypothesis class $\mF_{\rm \varphi}$ in eq.~\eqref{main:glm_hypothesis_class}. We focus on data ${\{\vec{x}_1^\intercal, \dots, \vec{x}_m^\intercal \} = \mat{X}\in \bbR^{m \times d}}$ drawn \iid from a distribution $P_x(\vec{x}) =\mN_{\vec{x}}(\vec{0},\mat{I}_d)$, and labels $\vec{y}$ drawn randomly from $ P_{y}(.)$. We consider for the moment a generic prior distribution $\vec{w} \sim P_w(.)$ that factorizes, and a component-wise activation function $\varphi(.)$. Let us define the cost function of a given sample $V(y_\mu|z_\mu) = \id\[ y_\mu \ne \varphi (z_\mu) \]$ that is 0 if the the estimator classifies the example correctly and 1 otherwise, where $z_\mu \equiv \frac{1}{\sqrt{d}} \vec{w}^{\intercal} \vec{x}_\mu$. Finally we define the constraint function at inverse temperature $\beta$, that depends explicitly on the Hamiltonian eq.~\eqref{main:hamiltonian}
\begin{equation}
    \mathcal{I}(\vec{y} | \vec{z}, \beta) \equiv \prod_{\mu=1}^m e^{- \beta V(y_\mu|z_\mu)} = e^{-\beta 	\mH\(\{\vec{y},\mat{X}\}, \vec{w}\) }\,,
    \label{appendix:replica_constraint}
\end{equation}
and note that the constraint function converges at zero temperature to a hard constraint function $ \mathcal{I}(\vec{y} | \vec{z}, \beta) \underset{\beta \to \infty}{\longrightarrow} \prod_{\mu=1}^m \id\[ V(y_\mu|z_\mu) = 0\]$. In order to compute the quenched free energy  average, we consider the partition function of $r \in \mathbb{N}$ identical copies of the initial system. We use the replica trick eq.~\eqref{main:replica_trick}. Assuming there exists an analytical continuation for $r \rightarrow 0$ and we can revert limits, the averaged free energy $\Phi$ of the initial system becomes eq.~\eqref{main:free_energy_trick}
\begin{align}
    \Phi(\alpha, \beta) = - \lim_{r \to 0} \[ \lim_{d \to \infty}  \frac{1}{d \beta}  \frac{\partial  \log \EE_{\vec{y}, \mat{X}} \[ \mZ(\{\vec{y}, \mat{X}\}, \alpha, \beta)^r\] }{\partial r} \] \,,
    \label{appendix:free_entropy}
\end{align}
where the replicated partition function reads using eq.~\eqref{main:partition_function}
\begin{align}
\begin{aligned}
&\EE_{\vec{y}, \mat{X}} \[ \mZ(\{\vec{y}, \mat{X}\}, \alpha, \beta)^r\] =  \int_{\bbR^m} \diff P_y\(\vec{y}\)  \int_{\bbR^{m\times d}} \diff P_x(\mat{X}) \mZ(\{\vec{y}, \mat{X}\}, \alpha, \beta)^r \\
&= \int_{\bbR^m} \diff P_y\(\vec{y}\)  \int_{\bbR^{m\times d}} \diff P_x(\mat{X})   \prod_{a=1}^r \int_{\bbR^d} \diff P_w\( \vec{w}^a \) \prod_{\mu=1}^m \int \d z_\mu^a  \mathcal{I}(y_\mu|z_\mu^a,\beta)\delta\left(z_\mu^a-\ \frac{1}{\sqrt{d}} \vec{w}^{a \intercal} \vec{x}_\mu \right)  \,.
\label{appendix:average_Zn}
\end{aligned}
\end{align}

\subsubsection{Average over $\mat{X}$ for \iid data}
As the data matrix is taken (Gaussian) \iid, for $i,j\in \llbracket 1;d \rrbracket$, $\mu,\nu \in \llbracket 1; m\rrbracket$, ${\EE_\mat{X}[x_{\mu i} x_{\nu j}] = \delta_{\mu\nu} \delta_{ij}}$. Hence $z_{\mu}^a =\frac{1}{\sqrt{d}} \sum_{i=1}^d x_{\mu i} w_i^a$ is the sum of \iid random variables. The central limit theorem guarantees that in the large size limit $d\to \infty$, $z_{\mu}^a \sim \mathcal{N}\left(\EE_{\mat{X}}[z_\mu^a]  ,\EE_{\mat{X}}[z_\mu^a z_\mu^b] \right)$, with the two first moments given by
\begin{equation}
	\begin{cases}
		\EE_{\mat{X}}[z_\mu^a] = \frac{1}{\sqrt{d}} \sum_{i=1}^d \EE_{\mat{X}}[x_{\mu i}] w_i^a =0\spacecase
		\EE_{\mat{X}}[z_\mu^a z_\nu^b] = \frac{1}{d} \sum_{ij} \EE_{\mat{X}}[x_{\mu i}x_{\nu j}] w_i^a w_j^b  = \frac{1}{d}\sum_{ij}  \delta_{\mu \nu} \delta_{ij} w_i^a w_j^b= \(\frac{1}{d} \sum_{i=1}^d w_i^a w_i^b \) \delta_{\mu \nu}  \, .
	\end{cases}
\end{equation}

In the following, we introduce the overlap matrix of size $r \times r$: $\mat{Q}\equiv\(\frac{1}{d}\vec{w}^a \cdot \vec{w}^b\)_{a,b=1..r}$ and we define $\td{\vec{z}}_{\mu} \in \bbR^{r} \equiv (z^a_{\mu})_{a=1..r}$, $\td{\vec{w}}_i \equiv (w_i^a)_{a=1..r} \in \bbR^{r}$. From the above calculation, $\td{\vec{z}}_{\mu}$ follows a multivariate Gaussian distribution $\td{\vec{z}}_{\mu} \sim P_{\td{z}}(\td{\vec{z}},\mat{Q}) \triangleq \mathcal{N}_{\td{\vec{z}}}( \tbf{0}_r, \mat{Q})$ and $
        P_{\td{w}}(\td{\vec{w}}_i) = \prod_{a=1}^r
        P_w(\td{w}^a_i)$. Introducing the change of variable and the Fourier representation of the $\delta$-Dirac function that involves a new matrix of size $r \times r$, $\hat{\mat{Q}}$:
\begin{align}
    1 &= \int_{\bbR^{r \times r}} \diff \mat{Q} \prod_{a \leq b} \delta \left(d Q_{ab}-\sum_{i=1}^d w_i^a     w_i^b \right) \\
    &\propto \int_{\bbR^{r \times r}} \diff \mat{Q} \int_{\bbR^{r \times r}} d\hat{\mat{Q}} \exp \left(-\frac{d}{2}\tr{\mat{Q} \hat{\mat{Q}}} \right)  \exp\left( \frac{1}{2}\sum_{i=1}^d \td{\vec{w}}_i^{\intercal} \hat{\mat{Q}} \td{\vec{w}}_i\right),
\end{align}
the replicated partition function factorizes and becomes an integral over the matrix parameters $\mat{Q}$ and $\hat{\mat{Q}}$, that can be evaluated using a Laplace method in the $d \to \infty$ limit,
\begin{align}
	\EE_{\vec{y}, \mat{X}} \[ \mZ(\{\vec{y}, \mat{X}\}, \alpha, \beta)^r\] &\propto \int \diff \mat{Q} \diff \hat{\mat{Q}} e^{d \Phi^{(r)}\(\mat{Q}, \hat{\mat{Q}}, \alpha, \beta \) } \underset{d \to \infty}{\simeq} e^{d \cdot  \extr_{\mat{Q}, \hat{\mat{Q}}} \left\{ \Phi^{(r)}\(\mat{Q}, \hat{\mat{Q}}, \alpha, \beta \) \right\}}, 
	\label{appendix:expectation_Zn}
\end{align}
where
\begin{equation}
     \begin{cases}
     \Phi^{(r)}\(\mat{Q}, \hat{\mat{Q}}, \alpha, \beta \) \equiv  -\frac{1}{2}\tr{\mat{Q}\hat{\mat{Q}}} +\log \Psi_{\rm w}^{(r)} (\hat{\mat{Q}})+\alpha\log \Psi_{\rm out}^{(r)}(\mat{Q},\beta)\,,
      \spacecase
      \Psi_{\rm w}^{(r)} (\hat{\mat{Q}}) = \displaystyle \int_{\mathbb{R}^r} \diff P_{\td{w}}(\td{\vec{w}})  e^{ \frac{1}{2}\td{\vec{w}}^{\intercal} \hat{\mat{Q}} \td{\vec{w}} }  \,, \spacecase
     \Psi_{\rm out}^{(r)}(\mat{Q},\beta) =  \displaystyle \int \diff P_y\(y\) \int_{\mathbb{R}^r}  \diff P_{\td{z}}(\td{\vec{z}},\mat{Q}) \mathcal{I}(y|\td{\vec{z}},\beta).
     \end{cases}
     \label{appendix:S_r}
\end{equation}
Finally, using eq.~\eqref{appendix:free_entropy} and switching the two limits $r \to 0$ and $d \to \infty$, the quenched free energy $\Phi$ simplifies as an extremization problem
\begin{equation}
\Phi (\alpha, \beta) = - \frac{1}{\beta} \extr_{\mat{Q}, \hat{\mat{Q}}} \left\{\lim_{r\rightarrow 0} \frac{\partial \Phi^{(r)} (\mat{Q},\hat{\mat{Q}}, \alpha, \beta)}{\partial  r} \right\} ,
\label{appendix:free_entropy2}
\end{equation}
over general symmetric matrices $\mat{Q}$ and $\hat{\mat{Q}}$. In the following we will assume simple ansatz for these matrices that allow to obtain analytic expressions in $r$ in order to take the derivative.

\paragraph{Choosing an ansatz}
Optimizing over the space of matrices is intractable, therefore one needs to assume a given ansatz about the matrices structure to push the computation further. The simplest and commonly used ansatz are the so-called
\begin{itemize}
    \item Replica Symmetry (RS) ansatz: $\mat{Q}^{(\rm rs)} = (Q-q_0) \mat{I}_r + q_0 \mat{J}_r$
    \item 1-Step Replica Symmetry Breaking (1RSB) ansatz: $\mat{Q}^{(\rm 1rsb)}= (Q-q_1) \mat{I}_r  + (q_1-q_0) \mat{I}_{r/x_0} \otimes \mat{J}_{x_0} + q_0 \mat{J}_r $\,,
    \item 2-Step Replica Symmetry Breaking (2RSB) ansatz: $\mat{Q}^{(\rm 2rsb)} = \left( Q - q_2 \right) \mat{I}_r + \left( q_2 - q_1 \right) \mat{I}_{r/x_1} \otimes \mat{J}_{x_1} +   \left( q_1 - q_0 \right) \mat{I}_{r/x_0} \otimes \mat{J}_{x_0} +q_0 \mat{J}_r  $
\end{itemize}
where $\mat{I}_k$ is the identity matrix of size $k$, and $\mat{J}_k$ is the matrix of size $k$ full of ones.
Plugging these ansatz and taking the derivative and the $r\to 0$ limit, optimizing over the space of matrices will boil down to a much simpler optimization problem over a few scalar order parameters. 

\subsubsection{RS free energy for \iid matrix}
\label{appendix:replica:rs}
%%%%%%%%%%%%%%%%%%%%%
Let us compute the functional $\Phi^{(r)} (\mat{Q},\hat{\mat{Q}}, \alpha, \beta)$ appearing in the free energy eq.~\eqref{appendix:free_entropy2} in the RS ansatz. The latter assumes that all replica remain equivalent with a common overlap $q_0 = \frac{1}{d} \sum_{i=1}^d w_i^a w_i^b$ for $a\ne b$ and a norm $Q= \frac{1}{d} \sum_{i=1}^d w_i^a w_i^a$, leading to the following expressions for matrices $\mat{Q}$ and $\hat{\mat{Q}} \in \mathbb{R}^{r\times r}$:
\begin{equation}
\begin{aligned}[c]
\mat{Q}^{(\rm rs)} =\begin{pmatrix} 
 Q & q_0 & ... & q_0 \\
 q_0 & Q & ... & ...  \\
 ... &... & ... & q_0  \\
 q_0 &... & q_0 & Q    \\
\end{pmatrix}
\end{aligned}
\hspace{0.5cm}
\textrm{ and } 
\hspace{0.5cm}
\begin{aligned}[c]
\hat{\mat{Q}}^{(\rm rs)}=\begin{pmatrix} 
 \hat{Q} & \hat{q}_0 & ... & \hat{q}_0\\
\hat{q}_0 &\hat{Q} & ... & ...  \\
 ... &... & ... & \hat{q}_0  \\
\hat{q}_0 &... & \hat{q}_0 & \hat{Q}\\  
\end{pmatrix} .
\end{aligned}
\end{equation}
Let us compute separately the terms involved in the functional $\Phi^{(r)} (\mat{Q},\hat{\mat{Q}}, \alpha, \beta)$ eq.~\eqref{appendix:S_r}: the first is a trace term, the second a term of prior $\Psi_{\rm w}^{(r)}$ and finally the third a term depending on the constraint $\mI(y|z)$ in eq.~\eqref{appendix:replica_constraint} $\Psi_{\rm out}^{(r)}$. 
\paragraph{Trace} 
The trace term can be easily computed and takes the form
\begin{equation}
	\left.\frac{1}{2}\tr{\mat{Q}\hat{\mat{Q}}} \right|_{\rm rs} =\frac{1}{2} \left( r Q\hat{Q} + r(r-1) q_0\hat{q}_0 \right)\,.
\end{equation}

\paragraph{Prior integral} Evaluated at the RS fixed point, and using a Gaussian identity also known as a Hubbard-Stratonovich transform, the prior integral can be further simplified 
\begin{align}
	\left.\Psi_{\rm w}^{(r)} (\hat{\mat{Q}})\right|_{\rm rs} &= \int \diff P_{\td{w}}(\td{\vec{w}})  e^{ \frac{1}{2}\td{\vec{w}}^{\intercal} \hat{\mat{Q}}_{\rm rs} \td{\vec{w}} } =   \int \diff P_{\td{w}}(\td{\vec{w}})  \exp{ \left( {\frac{(\hat{Q}- \hat{q}_0  )}{2}\sum_{a=1}^r (\td{w}^a)^2}\right)} \exp{ \left(\hat{q}_0 \left(  \sum_{a=1}^r  \td{w}^a  \right)^2 \right)} \nonumber \\
	&= \int \Diff \xi_0 \left [ \int \diff P_w(w)  \exp{ \left( {\frac{(\hat{Q}- \hat{q}_0  )}{2}  w^2}+ \xi_0 \sqrt{\hat{q}_0} w \right)}  \right]^r . 
\end{align}

\paragraph{Constraint integral} 
Recall the vector $\td{\vec{z}} \sim P_{\td{z}}(\td{\vec{z}},\mat{Q}) \triangleq \mathcal{N}_{\td{\vec{z}}}( \tbf{0}_r, \mat{Q})$ follows a Gaussian distribution with zero mean and covariance matrix $\mat{Q}$. In the RS ansatz, the covariance can be rewritten as a linear combination of the identity $\mat{I}_r$ and $\mat{J}_r$: $Q^{(\rm 2rsb)} = (Q-q_0) \mat{I}_r + q_0 \mat{J}_r$, that allows to split the variable $z^a = \sqrt{q_0} \xi_0 + \sqrt{Q-q_0} u^a  $ with  $\xi_0 \sim \mathcal{N}(0,1)$ and $ \forall a,~ u_a \sim \mathcal{N}(0,1)$.  The constraint integral then reads:
\begin{align}
\left. \Psi_{\rm out}^{(r)}(\mat{Q},\beta) \right|_{\rm rs} &= \displaystyle \int \diff P_y\(y\) \int_{\mathbb{R}^r}  \diff P_{\td{z}}(\td{\vec{z}},\mat{Q}) \mathcal{I}(y|\td{\vec{z}},\beta) \nonumber\\
& = \int \diff P_y\(y\) \int \Diff \xi_0 \int  \prod_{a=1}^r \Diff u^a  \mathcal{I}\(y|\sqrt{q_0} \xi_0 + \sqrt{Q-q_0} u^a , \beta \)\\
	 &= \int \diff P_y\(y\) \int \Diff \xi_0    \left[ \int  \Diff z  \mathcal{I}\(y|\sqrt{q_0} \xi_0 + \sqrt{Q-q_0} z , \beta \) \right]^r \,. \nonumber
\end{align}

Finally, putting pieces together, the functional $\Phi^{(r)} (\mat{Q},\hat{\mat{Q}}, \alpha, \beta)$ taken at the RS fixed point has an explicit formula and dependency in $r$:
\begin{align}
	\left.\Phi^{(r)} (\mat{Q},\hat{\mat{Q}}, \alpha, \beta) \right|_{\rm rs} &\underset{r\to 0}{\simeq} -\frac{1}{2} \left( r Q\hat{Q} + r(r-1) q_0\hat{q}_0 \right) \nonumber \\
	& + r \int \Diff \xi_0 \log\left( \int \diff P_w(w)  \exp{ \left( {\frac{(\hat{Q}- \hat{q}_0  )}{2}  w^2}+ \xi_0\sqrt{\hat{q}_0} w \right)}   \right) \\
	& + r\alpha  \int \diff P_y\(y\) \int \Diff \xi_0  \log\left(   \int  \Diff z  \mathcal{I}\(y|\sqrt{q_0} \xi_0 + \sqrt{Q-q_0} z , \beta \)  \right). \nonumber
\end{align}

\paragraph{Summary of RS free energy}
Taking the derivative with respect to $r$ and the $r\to 0$ limit, the RS free energy has a simple expression
\begin{align}\label{appendix:free_energy_rs}
    \Phi^{(\rm rs)}(\alpha, \beta) &= -\frac{1}{\beta} \underset{ q_0, \hat{q}_0}{\textbf{extr}}     \left\{   -\frac{1}{2}Q\hat{Q} + \frac{1}{2}q_0\hat{q}_0+  \Psi_{\rm w}^{(\rm rs)}(\hat{q}_0)   +\alpha  \Psi_{\rm out}^{(\rm rs)}(q_0, \beta)    \right\}, \\
    \Psi_{w\rm }^{\rm (rs)}(\hat{q}_0) &\equiv \EE_{\xi_0}  \log   \EE_{w} \[ \exp \( {\frac{(\hat{Q}- \hat{q}_0  )}{2}  w^2}+ \xi_0 \sqrt{\hat{q}_0} w \)  \]  \,, \\
	\Psi_{\rm out}^{\rm (rs)}(q_0, \beta) &\equiv  \EE_y \EE_{\xi_0}  \log  \EE_{z} \[\mI \(y \big | \sqrt{Q - q_0} z + \sqrt{q_0}\xi_0, \beta \)  \]\,, \label{appendix:replicas_out_rs}
\end{align} 
where $\xi_0,z \sim \mN(0,1)$, $w\sim P_w(.), y\sim P_y(.) $ and $Q=\hat{Q}=1$ in the case where $\frac{1}{d}\|\vec{w}\|_2^2=1$\\

\paragraph{Simplification in the spherical case}
In the spherical/Gaussian case, $\Psi_{\rm w}^{(r)} (\hat{\mat{Q}})$ in eq.~\eqref{appendix:S_r} can be directly integrated as
\begin{align}
	\Psi_{\rm w}^{(r)} (\hat{\mat{Q}}) = \displaystyle \int_{\|\td{\vec{w}}\|_2^2=d} d\td{\vec{w}}  e^{ \frac{1}{2}\td{\vec{w}}^{\intercal} \hat{\mat{Q}} \td{\vec{w}} } = -\frac{1}{2} \log \det \(2\pi (\id + \hat{\mat{Q}}) \) 
\end{align}
Besides, taking the derivative of eq.~\eqref{appendix:S_r} with respect to $\hat{\mat{Q}}$ we then find $\mat{Q}^{-1} = (\mat{I}_r + \hat{\mat{Q}})$. Finally we get rid of $\hat{\mat{Q}}$
\begin{align}
    \Phi^{(r)}\(\mat{Q}, \alpha, \beta \) &\equiv  \frac{1}{2} \log \det \(2\pi \mat{Q} \) +\alpha\log \Psi_{\rm out}^{(r)}(\mat{Q},\beta)\,.
	\label{app:replicas:spherical_simplification}	
\end{align}
\paragraph{Determinant}
The above determinant reads in the RS ansatz
\begin{equation}
	\left.\frac{1}{2}\det(\mat{Q}) \right|_{\rm rs} \simeq \frac{r}{2} \( \log(1-q_0) + \frac{q_0}{1 + (r-1)q_0} + \dots \)\,.
	\label{app:replicas:rs:determinant}
\end{equation}

Finally it leads in the RS ansatz to 
\begin{align}
\label{app:spherical:rs_free_energy_simplified}
	\Phi^{(\rm rs)}(\alpha,\beta) &= - \frac{1}{\beta} \extr_{q_0} \left\{ \frac{1}{2(1-q_0)} +\frac{1}{2}\log(2\pi) +\frac{1}{2} \log(1-q_0) +\alpha \Psi_{\rm out}^{\rm (rs)}(q_0, \beta) \right\}\,,
\end{align}
with $\Psi_{\rm out}^{\rm (rs)}$ defined in eq.~\eqref{appendix:replicas_out_rs}.

\subsubsection{1RSB free energy for \iid matrix}
\label{appendix:replica:1rsb}
The free entropy eq.~\eqref{appendix:free_entropy2} can also be evaluated at the simplest non trivial fixed point: the one-step Replica Symmetry Breaking ansatz (1RSB). Instead of assuming that replicas are equivalent, it states that the symmetry between replica is broken and that replicas are clustered in different states, with inner overlap $q_1$ and outer overlap $q_0$. Translating this analytically, the matrices can be expressed as function of the Parisi parameter $x_0$:
\begin{align}
\begin{aligned}
		\mat{Q}^{(\rm 1rsb)} &= q_0 \mat{J}_r + \left( q_1 - q_0 \right) \mat{I}_{\frac{r}{x_0}} \otimes \mat{J}_{x_0} +  \left( Q - q_1 \right) \mat{I}_r  \\
		\hat{\mat{Q}}^{(\rm 1rsb)} &= \hat{q}_0 \mat{J}_r + \left( \hat{q}_1 - \hat{q}_0 \right) \mat{I}_{\frac{r}{x_0}} \otimes \mat{J}_{x_0} +  \left( \hat{Q} - \hat{q}_1 \right) \mat{I}_r \,.
\end{aligned}
\end{align}

\paragraph{Trace} 
Again, the trace term can be easily computed
\begin{equation}
	\left.\frac{1}{2}\tr{\mat{Q}\hat{\mat{Q}}} \right|_{\rm 1rsb} =\frac{1}{2} \left( r Q\hat{Q} + r(x_0-1)q_1\hat{q}_1 + r(r-x_0)q_0\hat{q}_0 \right).
	\label{appendix:1RSB_Tr}
\end{equation}

\paragraph{Prior integral}
Separating replicas with different overlaps $q_0, q_1$, the prior integral can be written, using Hubbard-Stratonovich transformations to decouple replicas, as 
\begin{align}
	\left.\Psi_{\rm w}^{(r)} (\hat{\mat{Q}})\right|_{\rm 1rsb} &=  \int \diff P_{\td{w}}(\td{\vec{w}})  e^{  \frac{(\hat{Q}- \hat{q}_1  )}{2}\sum_{a=1}^r (\td{w}^a)^2 + \frac{(\hat{q}_1-\hat{q}_0)}{2} \sum_{k=1}^{\frac{r}{x_0}} \sum_{a,b=(k-1)x_0 + 1 }^{k x_0} \td{w}^a \td{w}^b + \frac{\hat{q}_0}{2} \left(\sum_{a=1}^r \td{w}^a \right)^2  } \nonumber\\
	&= \int \Diff \xi_0 \left[\int \Diff \xi_1\left[ \int \diff P_w(w) \exp\left( \frac{(\hat{Q}- \hat{q}_1  )}{2} w^2 + \left(\sqrt{\hat{q}_0}\xi_0+\sqrt{\hat{q}_1-\hat{q}_0}\xi_1 \right)w \right) \right ]^{x} \right]^{\frac{r}{x_0}}\,,
	\label{appendix:1RSB_Iw}
\end{align}
with $\xi_0, \xi_1 \sim \mN(0,1)$.

\paragraph{Constraint integral} Again, the vector $\td{\vec{z}} \sim P_{\td{z}} \triangleq \mN\(\tbf{0}, \mat{Q}^{(\rm 1rsb)}\)$  follows a gaussian vector with zero mean and covariance $ \mat{Q}^{(\rm 1rsb)} = q_0 \mat{J}_r + \left( q_1 - q_0 \right) \mat{I}_{\frac{r}{x_0}} \otimes \mat{J}_{x_0} +  \left( Q - q_1 \right) \mat{I}_r $. The gaussian vector of covariance $\mat{Q}^{(\rm 1rsb)}$ can be decomposed in a sum of normal gaussian vectors $\xi_0 \sim \mathcal{N}(0,1)$, ${\forall k \in \llbracket 1 ; \frac{r}{x_0}\rrbracket, \xi_k \sim \mathcal{N}(0,1)}$ and $ \forall a \in \llbracket (k-1) x_0 +1 ; k x_0 \rrbracket$, $u_a \sim \mathcal{N}(0,1)$:
\begin{equation*}
z^a = \sqrt{q_0} t_0 + \sqrt{q_1-q_0} t_k  + \sqrt{Q-q_1} u_{a}\,.
\end{equation*}
Finally the constraint integral reads
\begin{align}
&\left. \Psi_{\rm out}^{(r)}(Q,\beta) \right|_{\rm 1rsb} \nonumber\\ 
& = \int \diff P_y\(y\) \int \Diff \xi_0  \int \prod_{k=1}^{\frac{r}{x}}  \Diff \xi_k \int \prod_{a=(k-1)x_0+1}^{kx_0}  \Diff u_a   \mathcal{I}\(y|\sqrt{q_0} \xi_0 + \sqrt{q_1-q_0} \xi_k  + \sqrt{Q-q_1} u_{a}, \beta\) \nonumber \\
&= \int \diff P_y\(y\) \int \Diff \xi_0  \left[ \int \Diff \xi_1  \left[ \int \Diff z   \mathcal{I}\(y|\sqrt{q_0} \xi_0 + \sqrt{q_1-q_0} \xi_1  + \sqrt{Q-q_1} z, \beta \)\right]^{x_0} \right]^{\frac{r}{x_0}}.
\label{appendix:1RSB_Iz}
\end{align}
 
Gathering the previous computations eq.~(\ref{appendix:1RSB_Tr}, \ref{appendix:1RSB_Iw}, \ref{appendix:1RSB_Iz}), the functional $\Phi^{(r)}$ evaluated at the 1RSB fixed point reads:
\begin{align}
	&	\left.\Phi^{(r)} (\mat{Q},\hat{\mat{Q}}, \alpha, \beta) \right|_{\rm 1rsb} \nonumber\\
	&\underset{r\to 0}{\simeq}
	- \frac{1}{2} \left( r Q\hat{Q} + r(x_0-1)q_1\hat{q}_1 + r(r-x_0)q_0\hat{q}_0 \right) \\
	& + \frac{r}{x_0} \int \Diff \xi_0 \log\left(\int \Diff \xi_1\left[ \int \diff P_w(w) \exp\left( \frac{(\hat{Q}- \hat{q}_1  )}{2} w^2 + \left(\sqrt{\hat{q}_0}\xi_0+\sqrt{\hat{q}_1-\hat{q}_0}\xi_1 \right)w \right) \right ]^{x_0} \right) \nonumber \\
	& + \alpha\frac{r}{x_0} \int \diff P_y\(y\) \int \Diff \xi_0  \log\left(  \int \Diff \xi_1  \left[ \int \Diff z   \mathcal{I}\(y|\sqrt{q_0} \xi_0 + \sqrt{q_1-q_0} \xi_1  + \sqrt{Q-q_1} z, \beta \) \right]^{x_0}  \right). \nonumber
\end{align}

\paragraph{Summary and 1RSB free energy}
The free energy for the 1RSB ansatz is similar but more complicated and can be written as
\begin{multline}
    \label{appendix:free_energy_1rsb}
	\Phi^{(\rm 1rsb)}(\alpha, \beta) =  - \frac{1}{\beta}\underset{ \vec{q}, \hat{\vec{q}}, x_0}{\textbf{extr}} \left\{  \frac{1}{2} \left(  q_1\hat{q}_1 - Q\hat{Q} \right) + \frac{x_0}{2} \left(q_0\hat{q}_0 - q_1\hat{q}_1 \right)   \right. \\
	\left. + \Psi_{\rm w}^{(\rm 1rsb)}(\hat{\vec{q}})   +\alpha \Psi_{\rm out}^{(\rm 1rsb)}(\vec{q},\beta)    \right\} \,,
\end{multline}
\begin{align}
\begin{aligned}
	\Psi_{\rm w}^{(\rm 1rsb)}(\hat{\vec{q}}) &\equiv \frac{1}{x_0}
                \EE_{\xi_0} \log\left( \EE_{\xi_1} \EE_w \[\exp\left(
                  \frac{(\hat{Q}- \hat{q}_1  )}{2} w^2 +
                  \left(\sqrt{\hat{q}_0}\xi_0+\sqrt{\hat{q}_1-\hat{q}_0}\xi_1
                  \right)w \right)  \]^{x_0}\right) \,,  \\
    \Psi_{\rm out}^{(\rm 1rsb)}(\vec{q},\beta) &\equiv \frac{1}{x_0}
                \EE_y \EE_{\xi_0}  \log\left(  \EE_{\xi_1} \EE_{z} \[\mI( y \big | \sqrt{q_0} \xi_0 +
                \sqrt{q_1-q_0} \xi_1  + \sqrt{Q-q_1} z,\beta) \]^{x_0}\right)\,,       
\end{aligned}
\end{align}
where $\vec{q} =(q_0, q_1)$, $\hat{\vec{q}} =(\hat{q}_0, \hat{q}_1)$, $\xi_0, \xi_1,z \sim \mN(0,1)$, $w \sim P_w(.)$, $y\sim P_y(.)$ and $Q=\hat{Q}=1$.

\paragraph{Simplification in the spherical case}
The equation \eqref{app:replicas:spherical_simplification} remains valid. Let's compute the determinant in the 1RSB ansatz:
\paragraph{Determinant}
\begin{align}
	\left. \det \mat{Q}\right|_{\rm 1rsb} &= \( r q_0 + x_0 (q_1-q_0) +(1-q_1)\) \times \( 1 -q_1 \)^{r - r/x_0} \times \( x_0 (q_1-q_0) + (1-q_1) \)^{r/x_0-1}
\end{align}
Hence, 
\begin{align}
	\left. \log \det \mat{Q} \right|_{\rm 1rsb} \simeq r \( \frac{x_0-1}{x_0} \log (1- q_1) + \frac{1}{x_0} \log\( x_0(q_1-q_0) + (1-q_1) \) + \frac{q_0}{x_0(q_1-q_0) + (1-q_1)}  \)
	\label{app:det_1RSB}
\end{align}
Using the above expression for the determinant and the simplified replica potential in eq.~\eqref{app:replicas:spherical_simplification} we obtain
\begin{align}
	\Phi^{(\rm 1rsb)}(\alpha, \beta) = -\frac{1}{\beta} & \extr_{q_0, q_1, x_0} \left\{  \frac{1}{2}\log(2\pi) + \frac{x_0-1}{2x_0} \log (1- q_1) + \frac{1}{2x_0} \log\( x_0(q_1-q_0) + (1-q_1) \)  \right. \nonumber \\
	& \left. + \frac{q_0}{2 \( x_0(q_1-q_0) + (1-q_1) \) } +  \alpha   \Psi_{\rm out}^{(\rm 1rsb)}(\vec{q},\beta) \right\} \,.
	\label{app:spherical:1rsb_free_energy_simplified}
\end{align}

\subsubsection{2RSB free energy for \iid matrix}
Analogously, the 2RSB ansatz for $x_1 < x_0 < r$ is expressed by
\begin{align}
\begin{aligned}
		\mat{Q}^{(\rm 2rsb)} &= q_0 \mat{J}_r + \left( q_1 - q_0 \right) \mat{I}_{\frac{r}{x_0}} \otimes \mat{J}_{x_0} + \left( q_2 - q_1 \right) \mat{I}_{\frac{r}{x_1}} \otimes \mat{J}_{x_1}+  \left( Q - q_2 \right) \mat{I}_r  \\
		\hat{\mat{Q}}^{(\rm 2rsb)} &=  \hat{q}_0 \mat{J}_r + \left( \hat{q}_1 - \hat{q}_0 \right) \mat{I}_{\frac{r}{x_0}} \otimes \mat{J}_{x_0} + \left( \hat{q}_2 - \hat{q}_1 \right) \mat{I}_{\frac{r}{x_1}} \otimes \mat{J}_{x_1}+  \left( \hat{Q} - \hat{q}_2 \right) \mat{I}_r\,,
\end{aligned}
\end{align}
and the above computation of the free energy generalizes easily and yields
\begin{multline}
    \label{appendix:free_energy_2rsb}
	\Phi^{(\rm 2rsb)}(\alpha, \beta) =  - \frac{1}{\beta}\underset{ \vec{q}, \hat{\vec{q}}, x0, x_1}{\textbf{extr}} \left\{  \frac{1}{2} \left(  q_2\hat{q}_2 - Q\hat{Q} \right) + \frac{x_1}{2} \left(q_1\hat{q}_1 - q_2\hat{q}_2 \right)   + \frac{x_0}{2} \left(q_0\hat{q}_0 - q_1\hat{q}_1 \right)   \right. \\
	\left. + \Psi_{\rm w}^{(\rm 2rsb)}(\hat{\vec{q}})   +\alpha \Psi_{\rm out}^{(\rm 2rsb)}(\vec{q},\beta)    \right\} \,,
\end{multline}
\begin{align}
\begin{aligned}
    \Psi_{\rm w}^{(\rm 2rsb)}(\hat{\vec{q}}) &\equiv \frac{1}{x_1}
                \EE_{\xi_0} \log\left( \EE_{\xi_1} \[\EE_{\xi_2} \EE_w \[e^{
                  \frac{(\hat{Q}- \hat{q}_2  )}{2} w^2 +
                  \left(\sqrt{\hat{q}_0}\xi_0+\sqrt{\hat{q}_1-\hat{q}_0}\xi_1 + \sqrt{\hat{q}_2-\hat{q}_1}\xi_2
                  \right)w }  \]^{x_1} \]^{\frac{x_0}{x_1}} \right) \,,  \\
 	\Psi_{\rm out}^{(\rm 2rsb)}(\vec{q},\beta) &\equiv \frac{1}{x_1}
                \EE_y \EE_{\xi_0}  \log\left(  \EE_{\xi_1}\[ \EE_{\xi_2} \EE_{z} \[\mI( y \big | \sqrt{q_0} \xi_0 +
                \sqrt{q_1-q_0} \xi_1+ \sqrt{q_2-q_1} \xi_2  + \sqrt{Q-q_2} z,\beta)\]^{x_1} \]^{\frac{x_0}{x_1}} \right)\,.   
\end{aligned}
\end{align}

\subsubsection{Ground state energies - Spherical case}
\label{appendix:replicas_ground_state_spherical}
We focus on the particular case of the spherical perceptron with $\vec{w} \in \bbR^{d}$ on the sphere $\|\vec{w}\|_2^2=d$. 

\paragraph{RS ground state energy $e_{gs}$}
To compute the ground state energy, we first need to take both limits $q_0 \to 1$ and $\beta \to \infty$, keeping the product $\chi = \beta (Q-q_0)$ finite \cite{Majer1993a, Erichsen1992, Whyte1996}.
Recall eq.~\eqref{appendix:replicas_out_rs}
\begin{align}
	\Psi_{\rm out}^{\rm (rs)}(q_0, \beta) &\equiv  \EE_y \EE_{\xi_0}  \log  \EE_{z} \[\mI \(y \big | \sqrt{Q - q_0} z + \sqrt{q_0}\xi_0, \beta \)  \] \nonumber\\
	&= \int \diff P_y(y) \int \Diff \xi_0  \log\left(    \int  \mN_z\( \sqrt{q}_0 \xi_0 , Q - q_0\)  e^{-\beta V(y|z)  } \right)\\
	& \underset{(q_0,\beta) \to (1,\infty)}{\simeq}    -\frac{1}{2} \log(2\pi (Q-q_0)) - \beta \int \diff P_y(y) \int \Diff \xi_0 \min_{\xi, z} \[ V(y|z) +\frac{\(z - \xi_0 \)^2}{2 \chi}  \] \nonumber
	\label{app:spherical:gs:Psi_out}
\end{align}
that leads, taking limits $q_0 \to 1$, $\beta \to \infty$ in eq.~\eqref{app:spherical:rs_free_energy_simplified}, to the RS ground state energy
\begin{align}
    e_{\rm gs}^{\rm (rs)} &=  \extr_{\chi} \left\{ -\frac{1}{2 \chi} + \alpha  \EE_{y, \xi_0} \min_{z} \[ V(y|z) + \frac{\(z - \xi_0\)^2}{2 \chi}  \]   \right\}
\end{align}

\paragraph{Application to the step-perceptron}
Taking the step function $V(y|z) = \theta(\kappa-z)$ with ${P_y(y)=\delta(y-1)}$, it leads to the \cite{gardner1988optimal} expression:
\begin{align}
	 e_{\rm gs}^{\rm (rs)} &=  \extr_{\chi} \left\{ -\frac{1}{2\chi} + \alpha  \( \int_{-\infty}^{\kappa- \sqrt{2\chi}} \Diff \xi   + \int_{\kappa-\sqrt{2\chi}}^{\kappa} \Diff \xi \frac{(\xi - \kappa)^2}{2\chi} \)   \right\}
\end{align}

\paragraph{1RSB ground state energy $e_{gs}$}
\label{app:spherical:gs_1rsb}
To compute the ground state energy in the 1RSB ansatz, we first need to take limits $q_1 \to 1$ with $\beta \to \infty$ and $x_0\to 0$, keeping the products $ \chi \equiv \beta (Q-q_1)$  and $\omega_0 \equiv  x_0 \beta$ finite \cite{Whyte1996}, with $\Delta q = 1 - q_0$.
Recall 
\begin{align}
 &\Psi_{\rm out}^{(\rm 1rsb)}(\vec{q},\beta) \equiv \frac{1}{x_0}
                \EE_y \EE_{\xi_0}  \log\left(  \EE_{\xi_1} \EE_{z} \[\mI( y \big | \sqrt{q_0} \xi_0 +
                \sqrt{q_1-q_0} \xi_1  + \sqrt{Q-q_1} z,\beta) \]^{x_0} \right) \nonumber\\
     &= \frac{1}{x_0} \int \diff P_y(y)
                \int \Diff \xi_0  \log \int \Diff \xi_1   \( \int dz \mN_z \(\sqrt{q_0} \xi_0 +
                \sqrt{q_1-q_0} \xi_1 , 1-q_1  \) e^{-\beta V(y|z)} \)^{x_0} \\
     &\simeq \frac{1}{x_0}\int \diff P_y(y)
                \int \Diff \xi_0  \log \int \Diff \xi_1  e^{-x_0 \beta \min_z \[V(y|z) + \frac{1}{2 \beta (1-q_1)} \(z - \sqrt{q_0} \xi_0 -
                \sqrt{q_1-q_0} \xi_1 \)^2 \]}      \nonumber    
\end{align}

Finally, taking $q_1 \to 1$ with $\beta \to \infty$ and $x\to 0$ in eq.~\eqref{app:spherical:1rsb_free_energy_simplified}, defining  $\Omega_0 = \frac{\omega_0}{\chi}$, we obtain the 1RSB ground state energy
\begin{align}
e_{\rm gs}^{\rm (1rsb)} &= \extr_{\chi, \Omega_0, q_0} \left\{  \frac{1}{2\Omega_0 \chi} \log\( 1 + \Omega_0 \Delta q \) + \frac{q_0}{2\chi \( 1 + \Omega_0 \Delta q \) }   \right. \\
& \left. \hspace{4cm}  + \frac{\alpha}{\chi \Omega_0}  \EE_{\xi_0}  \log \EE_{\xi_1} e^{-\Omega_0 \chi \min_z \[V(y | z) + \frac{1}{2 \chi } \(z - \sqrt{q_0} \xi_0 -
                \sqrt{\Delta q} \xi_1 \)^2 \]}   \right\}\,. \nonumber
\end{align}

\paragraph{2RSB ground state energy $e_{gs}$}
\label{app:spherical:gs_2rsb}

Taking $q_2 \to 1$ with $\beta \to \infty$, we define $\Omega_0= \frac{x_0 \beta}{\chi}, \Omega_1= \frac{x_1 \beta}{\chi}$ and we obtain similarly the 2RSB ground state energy of the spherical perceptron:
\begin{align}
\begin{aligned}
	e_{\rm gs, iid}^{\rm (2rsb)} &= \extr_{\chi,\Omega_1, \Omega_0, q_1,q_0,} \left\{ \frac{q_0}{2\chi(1+\Omega_1 (1-q_1) + \Omega_0 (q_1-q_0)} +  \frac{1}{2 \Omega_1 \chi} \log(1+\Omega_1 (1-q_1))   \right. \\ 
	&\left. + \frac{1}{2\Omega_0\chi} \log\(1+\frac{\Omega_0(q_1-q_0)}{1+\Omega_1 (1-q_1)}\) \right.\\
	&\left. + \frac{\alpha}{\chi \Omega_0} \EE_{\xi_0} \log \EE_{\xi_1} \[\EE_{\xi_2} e^{-\Omega_1 \chi \min_z \[V(y | z) + \frac{1}{2 \chi } \(z - \sqrt{q_0} \xi_0 -
                \sqrt{q_1-q_0} \xi_1 - \sqrt{1-q_1} \xi_2 \)^2 \]}  \]^{\Omega_0 / \Omega_1} \right\}
\end{aligned}
\end{align}
and note that taking $q_1=q_0, x_0=x_1$ we recover the 1RSB expression.

\subsection{Rotationally invariant matrix}
\label{appendix:replicas_rot_inv_mat}
The replica free energy can be derived in the general setting when the data matrix $\mat{X}\in \bbR^{m \times d}$ is orthogonally invariant~\cite{Kabashima2008}. This includes Gaussian \iid matrices, but encompasses a larger class of matrices with correlation among data samples. In this section we provide expressions that could be useful for the interested reader, that might want to compute the ground state energy in this case.

\subsubsection{RS free energy for rotationally invariant matrices}
The replica symmetric free energy when $\mat{X}$ is rotationally invariant ($\rm{RI}$) reads
\begin{multline}
    \Phi^{\rm (rs)}_{\rm RI}(\alpha, \beta) = - \dfrac{1}{\beta} \extr_{\chi_w, \chi_u, q_w, q_u} \left\lbrace \mathcal{A}_0^{\rm (rs)} (\chi_w, \chi_u, q_w, q_u) \right.  \\
    \left. \hspace{3cm} + \mathcal{A}_w^{\rm (rs)} (\chi_w, q_w) + \alpha \mathcal{A}_u^{\rm (rs)} (\chi_u, q_u, \beta)\right\rbrace
    \label{appendix:free_energy_RI_rs}
\end{multline}
where
\begin{align}
\mathcal{A}_0^{\rm (rs)} (\chi_w, \chi_u, q_w, q_u) &\equiv F(\chi_w, \chi_u) + q_w \dfrac{\partial F (\chi_w, \chi_u)}{\partial \chi_w} - q_u \dfrac{\partial F (\chi_w, \chi_u)}{\partial \chi_u}\,,
\end{align}
\begin{multline}
    \mathcal{A}_w^{\rm (rs)}(\chi_w, q_w) \equiv \extr_{\hat{\chi}_w, \hat{q}_w} \left\lbrace\dfrac{\hat{\chi}_w}{2}(\chi_w + q_w) - \dfrac{\hat{q}_w}{2} \chi_w \right. \label{Aw_RS} \\
    \left.  + \int \Diff \xi_0 \log \left\lbrace \int \diff P_w(w) \exp \left[ -\dfrac{\hat{\chi}_w}{2}w^2 + \sqrt{\hat{q}_w}\xi_0 w\right] \right\rbrace \right\rbrace\,,
\end{multline}
\begin{multline}
    \mathcal{A}_u^{\rm (rs)}(\chi_u, q_u, \beta) \equiv \extr_{\hat{\chi}_u, \hat{q}_u} \left\lbrace\dfrac{\hat{\chi}_u}{2}(\chi_u - q_u) + \dfrac{\hat{q}_u}{2} \chi_u \right. \label{Au_RS} \\
    \left.  + \int \diff P_y(y) \int \Diff \xi_0 \log \left\lbrace \int \Diff z \mathcal{I} \left( y | \sqrt{\hat{\chi}_u} z + \sqrt{\hat{q}_u} \xi_0, \beta \right)\right\rbrace \right\rbrace \,,
\end{multline}
and given $\rho(\lambda)$ the supposedly well-defined asymptotic eigenvalue distribution of $\mat{X}^\intercal \mat{X}$,
\begin{align}
    F(x,y) &\equiv \extr_{\Lambda_x, \Lambda_y} \left\lbrace -\dfrac{1}{2}\langle \log (\Lambda_x \Lambda_y + \lambda)\rangle_\rho - \dfrac{\alpha -1}{2}\ln \Lambda_y + \dfrac{\Lambda_x x}{2} + \alpha\dfrac{\Lambda_y y}{2} \right\rbrace \nonumber \\
    & \hspace{3cm}- \dfrac{1}{2} \log x - \dfrac{\alpha}{2}\log y - \dfrac{1+\alpha}{2}.
\end{align}
Let us check if we recover the replica free energy for a Gaussian i.i.d. matrix eq.~\eqref{appendix:free_energy_rs}. In that case, the eigenvalue distribution $\rho(\lambda)$ follows the Marchenko-Pastur distribution, and we obtain $F_{\rm iid}(x,y) = - \dfrac{\alpha}{2}x y$. To match notations, we rename quantities $\hat{\chi}_w^\star$, $\hat{q}_w^\star$, $\hat{\chi}_u^\star$, $\hat{q}_u^\star$ (extremized values of hat parameters inside $\mathcal{A}_u$ and $\mathcal{A}_w$) following 
\begin{equation}
    \hat{\chi}_w \rightarrow \hat{q}_0 - \hat{Q}, \hspace{.5cm}\hat{q}_w \rightarrow \hat{q}_0, \hspace{.5cm}\hat{\chi}_u \rightarrow Q-q_0, \hspace{.5cm}\hat{q}_u \rightarrow q_0. 
\end{equation}
We take the derivative of $ \Phi^{\rm (rs)}_{\rm RI}$ with respect to $\chi_w$, $\chi_u$, $q_w$, $q_u$ to get
\begin{align}
\begin{cases}
    \alpha \chi_u &= \hat{q}_0 - \hat{Q}\\
    \chi_w &= Q - q_0 \\
    \alpha q_u &= \hat{q}_0 \\
    q_w &= q_0.
\end{cases}
\end{align}
Finally, in agreement with eq.~\eqref{appendix:free_energy_rs} we reach
\begin{multline}
     \Phi^{(\rm rs)}_{\rm iid}(\alpha, \beta) = -\dfrac{1}{\beta}\extr_{q_0, \hat{q}_0} \left\lbrace -\dfrac{1}{2}Q\hat{Q} + \frac{1}{2}q_0\hat{q}_0 + \right. \\
     \left. + \int \Diff \xi_0 \log \left\lbrace \int \diff w P_w(w) \exp \left[ \dfrac{\hat{Q}-\hat{q}_0}{2}w^2 + \xi_0 \sqrt{\hat{q}_0} w \right] \right\rbrace \right. \\ \left. + \alpha \int \diff P_y(y) \int \Diff \xi_0 \log \left\lbrace \int \Diff z \mathcal{I} \left( y | \sqrt{q_0} \xi_0 + \sqrt{Q-q_0}z, \beta \right) \right\rbrace \right\rbrace.
\end{multline}

\subsubsection{1RSB free energy for rotationally invariant matrices}
The 1RSB free energy for rotationally invariant matrices is also derived in~\cite{Kabashima2008} and reads
\begin{multline}
    \Phi^{\rm(1rsb)}_{\rm RI}= - \dfrac{1}{\beta}\extr_{\chi_w, \chi_u, v_w, v_u, q_w, q_u, m} \left\lbrace \mathcal{A}_0^{\rm (1rsb)} (\chi_w, \chi_u, v_w, v_u, q_w, q_u, m) +  \right. \\ 
    \left. + \mathcal{A}_w^{\rm (1rsb)}(\chi_w, v_w, q_w, m) + \alpha \mathcal{A}_u^{\rm (1rsb)}(\chi_u, v_u, q_u, m, \beta)\right\rbrace
\end{multline}
with
\begin{multline}
    \mathcal{A}_0^{\rm (1rsb)} (\chi_w, \chi_u, v_w, v_u,  q_w, q_u, m) \equiv F(\chi_w, \chi_u) + \dfrac{1}{m} \left[ F(\chi_w + m v_w, \chi_u - m v_u) - F (\chi_w, \chi_u) \right] \\ 
    + q_w \dfrac{\diff F (\chi_w + m v_w, \chi_u - m v_u)}{\diff \chi_w} - q_u \dfrac{\diff F (\chi_w + m v_w, \chi_u - m v_u)}{\diff \chi_u}
\end{multline}
\begin{multline}
    \mathcal{A}_w^{\rm (1rsb)} (\chi_w, \chi_w, v_w, m) \equiv \extr_{\hat{\chi}_w, \hat{v}_w, \hat{q}_w} \left\lbrace \dfrac{\hat{\chi}_w (\chi_w + v_w + q_w)}{2} - \dfrac{\hat{v}_w (\chi_w+m(v_w + q_w))}{2}  \right. \\ 
    \left. - \dfrac{\hat{q}_w (\chi_w + m v_w)}{2} + \dfrac{1}{m} \int \Diff z \log \left[ \int \Diff y \left( \text{Tr}_w P_w(w) e^{-\frac{\hat{\chi}_w}{2}w^2 + (\sqrt{\hat{v}_w y + \sqrt{q}_w z}) w} \right)^m \right] \right\rbrace
\end{multline}
\begin{multline}
    \mathcal{A}_u^{\rm (1rsb)}(\chi_u, v_u, q_u, m, \beta) \equiv \extr_{\hat{\chi}_u, \hat{v}_u, \hat{q}_u} \left\lbrace \dfrac{\hat{\chi}_u (\chi_u - v_u - q_u)}{2} + \dfrac{\hat{v}_u (\chi_u-m(v_u + q_u))}{2}  \right. \\  
     + \dfrac{\hat{q}_u (\chi_u - m v_u)}{2} + \left. \dfrac{1}{m}\int \diff y P_y(y) \int \Diff z \log \left[ \int \Diff s \left( \int \Diff x \mathcal{I} \left( y | \sqrt{\hat{\chi}_u}x + \sqrt{\hat{v}_u}s + \sqrt{\hat{q}_u} z, \beta \right)\right)^m \right] \right\rbrace.
\end{multline}
We rename the extremized hat variables $\hat{\chi}^\star_w, \hat{v}^\star_w, \hat{q}^\star_w, \hat{\chi}^\star_u, \hat{v}_u, \hat{q}_u$ as
\begin{equation}
    \hat{\chi}_w \rightarrow \hat{q}_1 - \hat{Q}, \hspace{.6cm} \hat{v}_w \rightarrow \hat{q}_1 - \hat{q}_0, \hspace{.6cm} \hat{q}_w \rightarrow \hat{q}_0, \hspace{.6cm} \hat{\chi}_u \rightarrow Q - q_1, \hspace{.6cm} \hat{v}_u \rightarrow q_1 - q_0, \hspace{.6cm} \hat{q}_u \rightarrow q_0.
\end{equation}
We explicit $\mathcal{A}_0$ using $F_{\iid} (x,y) = - \dfrac{\alpha}{2} x y$, then take the derivatives of $\Phi^{\rm (1rsb)}_{\rm RI}$ with respect to $\chi_w$, $\chi_u$, $q_w$, $q_u$, $v_w$, $v_u$. After some steps we obtain
\begin{align}
\begin{cases}
    \chi_w &= Q - q_1\\
    \alpha \chi_u &= \hat{q}_1 - \hat{Q} \\
    q_w &= q_0 \\
    \alpha q_u &= \hat{q}_0 \\
    v_w &= q_1 - q_0 \\
    \alpha v_u &= \hat{q}_1 - \hat{q}_0.
\end{cases}
\end{align}
Replacing all this in $\Phi^{\rm (1rsb)}_{\rm RI}$, we find
\begin{multline}
    \Phi^{\rm(1rsb)}_{\rm iid} (\alpha, \beta)= - \dfrac{1}{\beta}\extr_{q_0, q_1, \hat{q}_0, \hat{q}_1, x} \left\lbrace \dfrac{1}{2}(q_1 \hat{q}_1 - Q \hat{Q}) + \dfrac{x}{2}(q_0 \hat{q}_0 - q_1 \hat{q}_1) \right. \\+\dfrac{1}{x} \int \Diff \xi_0 \log \left\lbrace  \int \Diff \xi_1 \int \diff w P_w(w) \exp \left[ \dfrac{\hat{Q}-\hat{q}_1}{2} w^2 + \left( \sqrt{\hat{q}_0} \xi_0 + \sqrt{\hat{q}_1-\hat{q}_0}\xi_1\right)w\right]^x \right\rbrace  \\ 
    \left. + \dfrac{\alpha}{x} \int \diff P_y(y)\int \Diff \xi_0 \log \left\lbrace \int \Diff \xi_1 \left[ \int \Diff z \mathcal{I}\left( y | \sqrt{q_0}\xi_0 + \sqrt{q_1 - q_0}\xi_1 + \sqrt{Q-q_1}z, \beta \right)\right]^x\right\rbrace \right\rbrace
\end{multline}
in agreement with eq.~\eqref{appendix:free_energy_1rsb}.

\end{document}